\documentstyle[12pt,epsfig]{article}

\def\beq{\begin{equation}}
\def\eeq{\end{equation}}
\def\barr{\begin{eqnarray}}
\def\earr{\end{eqnarray}}
\def\lsim{\raise0.3ex\hbox{$\;<$\kern-0.75em\raise-1.1ex\hbox{$\sim\;$}}}
\def\gsim{\raise0.3ex\hbox{$\;>$\kern-0.75em\raise-1.1ex\hbox{$\sim\;$}}}
\def\lg{\raise0.3ex\hbox{$\;>$\kern-0.75em\raise-1.1ex\hbox{$<\;$}}}
\def\equiapp{\raise0.3ex\hbox{$\;\sim$\kern-0.75em\raise-1.1ex\hbox{$=\;$}}}

\begin{document}

\rightline{IC/99/83}
\rightline{hep-ph/9907423}

\bigskip

\begin{center}

{\Large \bf Identifying the neutrino mass spectrum from
a supernova neutrino burst}

\bigskip

{\it Amol S. Dighe \footnote{\tt amol.dighe@cern.ch}, 
Alexei Yu. Smirnov \footnote{\tt smirnov@ictp.trieste.it} \\ 
The Abdus Salam International Center for
Theoretical Physics \\ 34100, Trieste, Italy.}

\end{center}

\begin{abstract}

We study the role that the future
detection of the neutrino burst from a galactic supernova 
can play in the reconstruction of the neutrino mass spectrum.
We consider all possible 3$\nu$ mass and flavor spectra  
which describe the solar and atmospheric neutrino data. 
For each of these spectra  we 
find the observable effects of the supernova neutrino conversions  
both in the matter of the star and the earth.  
We show that studies of the electron neutrino and anineutrino spectra 
as well as observations of the neutral current effects from supernova will
allow us 
(i) to identify  the solar neutrino solution, 
(ii) to determine the type of mass hierarchy (normal or inverted) and 
(iii) to probe the mixing $|U_{e3}|^2$ to values as low as 
$10^{-4} - 10^{-3}$.

\end{abstract}

\newpage

\section{Introduction}
\label{intro}

The reconstruction of the neutrino mass and flavor spectrum is 
one of the fundamental problems of particle physics. It also has
important implications for cosmology and astrophysics. 
The knowledge of neutrino masses and mixing will
allow us to clarify the role of neutrinos in the mechanism of
the star explosion and supernova nucleosynthesis.

With the present data on the atmospheric and solar neutrinos, we  
are taking the  first steps in the reconstruction of the spectrum.
The SuperKamiokande (SK) 
results \cite{sk} on atmospheric neutrinos, confirmed
by the recent SOUDAN \cite{soudan} and MACRO \cite{macro} data,
allow us to claim with a high confidence level that the
atmospheric neutrinos  oscillate. Moreover,
the oscillations are due to neutrino masses and the mixing 
in vacuum. The data also indicates 
$\nu_\mu \leftrightarrow \nu_\tau$ as the dominant mode. All the 
existing experimental results can be well described in terms of
the $\nu_\mu \leftrightarrow \nu_\tau$ vacuum oscillations
with the mass squared difference and the mixing parameters given by
\cite{sk}
\beq
|\Delta m^2_{atm}| = (1 - 8) \cdot 10^{-3} \mbox{ eV}^2 ~~,~~
\sin^2 2\theta = 0.8 - 1.0~~.
\eeq
There is no compelling evidence that the electron neutrinos
participate in the oscillations of atmospheric neutrinos. Moreover,
the CHOOZ experiment \cite{chooz} gives an upper 
bound on the mixing of $\nu_e$ with $\Delta m^2 \sim \Delta
m^2_{atm}$:
\beq
\sin^2 2 \theta_e \leq 0.1 \quad \mbox{ for } \quad
|\Delta m^2| > 2 \cdot 10^{-3}  \mbox{ eV}^2~~.
\eeq

The oscillation interpretation of the atmospheric neutrino data
indicates that the solution of the solar 
neutrino problem is also related to nonzero neutrino masses
and mixing. At the moment, however, there are
several possible
solutions. Moreover, various
sorts of data --
spectral distortions, day-night effects, 
seasonal variations --
favor different possible solutions. 
A good desription of all the existing data can be 
obtained by~\cite{solar-fit}:
\begin{enumerate}
\item the small mixing angle (SMA) MSW solution:
\beq
\Delta m^2_\odot = (4 - 10) \cdot 10^{-6}   \mbox{ eV}^2~~,~~
\sin^2 2 \theta_\odot = (2 - 10) \cdot 10^{-3}~~,
\label{sma-param}
\eeq
\item the large mixing angle (LMA) MSW solution:
\beq
\Delta m^2_\odot = (1 - 10) \cdot 10^{-5}   \mbox{ eV}^2~~,~~
\sin^2 2 \theta_\odot = 0.7 - 0.95~~,
\label{lma-param}
\eeq 
\item the vacuum oscillation (VO) solution:
\beq
\Delta m^2_\odot =   
\begin{array}{r}
(4 - 6) \cdot 10^{-10}   \mbox{ eV}^2 \\
(6 - 8) \cdot 10^{-11}   \mbox{ eV}^2 
\end{array} ~~,~~
\sin^2 2 \theta_\odot = 0.8 - 1.0~~.
\label{vo-param}
\eeq 
\end{enumerate}
Some other possibilities are also not excluded --
{\it e.g.} the LOW MSW solution with 
$\Delta m^2 \sim (0.5 - 2) \cdot 10^{-7}$ eV$^2$ 
and $\sin^2 2 \theta_\odot = 0.9 - 1.0$
(see \cite{solar-fit,lowdmsq}). 
Results from future experiments with the existing and new detectors will
remove this ambiguity, thus identifying the correct solution to
the solar neutrino problem.

Another evidence for the neutrino oscillations follows from
the LSND results \cite{lsnd}, which are not confirmed, but also
not excluded by the KARMEN experiment \cite{karmen}. The LSND
results cannot be reconciled with the solutions of the atmospheric
and solar neutrino problems in the context of only three known 
neutrinos, thus requiring the introduction of sterile neutrinos.
\cite{ds-sterile}. 
In this paper, we shall consider only the mixing between the known
three neutrinos (spectra with sterile neutrinos  
will be discussed elsewhere).

The atmospheric and solar neutrino results 
(\cite{sk}-\cite{macro}, \cite{solar-fit}) as well as the existing 
bounds from
the other oscillation experiments and the $\beta \beta _{0 \nu}$
searches lead to several possible spectra of neutrino
masses and mixing. The ambiguity is
related to (i) the unidentified solution of the solar neutrino
problem (ii) unknown mixing of $\nu_e$ in the
third mass eigenstate which is described by the matrix element $U_{e3}$ 
(iii) type of hierarchy (normal or inverted) which is related to 
the mass of the third mass eigenstste 
(whether it is the lightest or the heaviest one). 
The absolute scale of mass is also unknown, however this 
cannot be established from the oscillation phenomena. \\

In this paper, we reconsider the effects of oscillations on
supernova neutrinos. 
With the existing data on neutrino masses and mixing,
we can sharpen the predictions of the oscillation effects in the
supernova neutrinos. On the other hand, we  
clarify the extent to which the studies of supernova neutrinos
can contribute to the reconstruction of the neutrino masses and flavor
spectrum. We will show that three ambiguities mentioned above 
can in principle be resolved by supernova data.

The effects of neutrino mixing on the neutrino fluxes
from the supernova have been extensively discussed
in the context of $2\nu$ mixing.
For a wide range of mixing parameters
($\Delta m^2 \lsim 10^4$ eV$^2$), the neutrinos encounter
their MSW resonance densities inside the star, hence
the studies of resonant neutrino conversions inside the star
\cite{sn-resonant}-\cite{notzold} are crucial.
For very low values of $\Delta m^2$ ({\it e.g.} 
$\Delta m^2 \lsim 10^{-14}$ eV$^2$), the
vacuum oscillations on the way from the star to the earth
need to be taken into account \cite{stodolsky,sn-vac}.
In the presence of a strong magnetic field,  
the spin-flip effects become important \cite{athar}:
the spin-flavor precession \cite{fujikawa} and 
resonant spin-flavor conversions \cite{akhmedov}
may affect the observed neutrino fluxes.
If sterile neutrinos are involved in the neutrino
conversions, they may enable the r-process nucleosynthesis
\cite{mclaugh}.

The effects of the neutrino conversions 
can be observed through, {\it e.g.}
(i) the disappearance (partial or complete) of the neutronization
peak;
(ii) the interchange of original spectra and the 
appearance of a hard $\nu_e$ spectrum;
(iii) distortions of the $\nu_e$ energy spectrum;
(iv) modification of the $\bar{\nu}_e$ spectrum
(in particular, the effects of large lepton mixing on the
$\bar{\nu}_e$ spectrum have been extensively studied
\cite{spergel});
(v) earth matter effects.
The observation of the neutrino burst from SN1987A \cite{sn87a}
has already given
bounds on the large mixing of active neutrinos
(\cite{sn87-resonant,notzold},\cite{spergel}-\cite{walker}) 
and on the mixing of
$\nu_e$ with sterile neutrinos \cite{e-sterile-bounds}.

The main features of transitions of supernova neutrinos
in the case of $3\nu$ mixing 
(\cite{walker},\cite{kuo-prd37}-\cite{dutta})
can be understood in terms of the $2\nu$ mixing. 
The system has two resonances\footnote{The
radiative corrections to $m_{\nu_\mu}$ and $m_{\nu_\tau}$
imply the existence of one more resonance between the two non-electron
neutrinos, but since the two non-electron neutrinos cannot be 
distinguished at the detector, the conversions between them
do not affect the observations. See sec.~\ref{levelx}.} 
Under the assumptions of 
mass hierarchy and smallness of mixing, the dynamics of
the two level crossings splits. As a result, the
factorization of probabilities occurs \cite{kuo-prd37,ms-ppnp}.
In the presence of sterile neutrinos, multi-level conversions
take place \cite{japan,multilevel-sterile}, which may be
interpreted in terms of the constituent $2\nu$ conversions
for small mixing angles and $\Delta m^2$ hierarchy
\cite{ds-sterile}.\\

In this paper, we study the conversions of supernova
neutrinos in the $3\nu$ context, taking into account
the recent results on the neutrino masses and mixing. We 
consider the effects for all  possible schemes of neutrino
masses and mixing which explain the atmospheric and solar 
neutrino data. For each scheme, we  find the modifications
of (i) the neutronization peak (ii) the $\nu_e$ 
energy spectrum and
(iii) the $\bar{\nu}_e$ energy spectrum,
which can be observed directly. We also determine 
the spectrum of the non-electron neutrinos which can in principle be
studied by neutral current interactions in reactions with 
different energy thresholds.

The paper is organized as follows. 
In Sec. II we describe the
features of initial neutrino fluxes from the supernova and the
dynamics of neutrino conversion on their way out to the 
surface of the star. 
In Sec. III, we  derive general expressions for the
transition probabilities for the schemes with normal mass
hierarchy. We also calculate the earth matter effects on
the neutrino  spectra.
In Sec. IV, 
we find the final neutrino spectra at detectors  
for the schemes with the normal mass hierarchy. 
In Sec. V, 
we perform similar studies  for the schemes with
the inverted mass hierarchy.
In Sec. VI, we discuss the observable signals 
and the signatures of various mixing schemes 
In Sec. VII, comparing results for various schemes we conclude about   
the possibility to discriminate the schemes by 
future observations of neutrino bursts from
a galactic supernova. 

%%%%%%%%%%%%%%%%%%%%%%%%%%%%%%%%%%%%%%%%%%%%%%%%%%%%%%%%%%%%%%%%%%%%%
%%%%%%%%%%%%%%%%%%%%%%%%%%%%%%%%%%%%%%%%%%%%%%%%%%%%%%%%%%%%%%%%%%%%%%
\section{Mass spectra, fluxes and dynamics of conversion}
\label{conv}
%%%%%%%%%%%%%%%%%%%%%%%%%%%%%%%%%%%%%%%%%%%%%%%%%%%%%%%%%%%%%%%%%%%%%%

In this section, the generic properties of the initial neutrino fluxes
will be summarized. 
We identify  the neutrino mass and mixing
parameters relevant for the supernova neutrino 
conversions, 
and consider main  aspects of dynamics of neutrino conversion inside
the star:  the transition regions, the factorization of
dynamics and adiabaticity. Finally, we  construct the level crossing
schemes for the normal and  the inverted mass hierarchy.\\

%%%%%%%%%%%%%%%%%%%%%%%%%%%%%%%%%%%%%%%%%%%%%%%%%%%%%%%%%%%%%%%%%%%%%%
\subsection{Neutrino fluxes}
\label{fluxes}
%%%%%%%%%%%%%%%%%%%%%%%%%%%%%%%%%%%%%%%%%%%%%%%%%%%%%%%%%%%%%%%%%%%%

In what follows we will  summarize the generic features of the 
original fluxes which do not depend on the model and
the parameters of the star. The   deviations  from these 
features will testify for neutrino conversions.

{\bf 1. Flavor of the neutronization peak.} 
During first few milliseconds of the neutrino burst
from a supernova, the signal
is expected to be dominated by the $\nu_e$,
which are produced by the electron capture on
protons and nuclei while the shock wave passes  through
the neutrinosphere \cite{neutronization}.
Since the original flux is $\nu_e$, the
final observed fluxes give a direct measurement of
the extent of conversion of $\nu_e$ into
the other neutrino species.\\

{\bf 2. Inequalities of average energies of the spectra.} 
Since $\nu_e$ interact more strongly with
matter than the other species, 
their effective neutrinosphere is outside
the neutrinospheres of the other species and hence they have
a lower average energy  than $\bar{\nu}_e$ and $\nu_x$.
The $\bar{\nu}_e$ also interact via charged current,
but the cross section is smaller, so their
average energy is more  than that of the $\nu_e$, but less
than that of $\nu_x$.

\beq
\langle E^0_{\nu_e} \rangle  <
\langle E^0_{\bar{\nu}_e} \rangle  <
\langle E^0_{\nu_x} \rangle~~.
\label{e-hierarchy}
\eeq
Here subscript $0$ refers to the original spectra. The neutrino conversion 
changes inequalities (\ref{e-hierarchy}). \\

{\bf 3. The pinched  spectra during the cooling stage.} 
Let $F_e^0$ and $F_{\bar{e}}^0$ be the original fluxes 
of $\nu_e$ and $\bar{\nu}_e$  respectively, produced at the cooling stage. 
The ``non-electron'' neutrinos ($\nu_\mu, \nu_\tau, 
\bar{\nu}_\mu, \bar{\nu}_\tau$) have the
same neutral current interactions inside the supernova,
and their
original fluxes are expected to be approximately equal\footnote{
The presence of real muons in the central part of the star 
leads to a nonzero chemical potential of the muon neutrinos 
and hence to a difference of fluxes \cite{horowitz}. 
However, in the neutrinosphere 
with $T \approx 6 - 8$ MeV, the concentration of muons is smaller than 
$1\%$.}. 
In what follows we will neglect the difference of fluxes.
We will denote these four neutrino species collectively as $\nu_x$ and
the original flux of each of them 
as  $F_x^0$. 
 
\begin{figure}[htb]
\hbox to \hsize{\hfil\epsfxsize=8cm\epsfbox{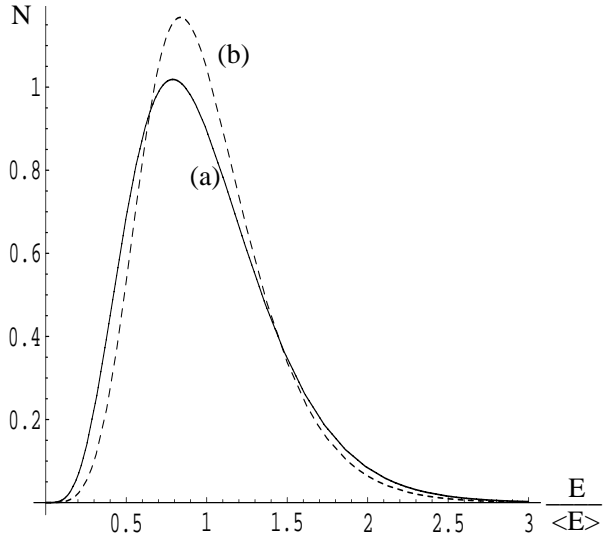}\hfil}
\caption{~~The number of $\nu-N$ charged current events for
(a) a thermal spectrum ($T=3, \eta = 0$)
(b) a ``pinched'' spectrum  ($T = 3, \eta = 3$).
$T$ and $\eta$ are the parameters as in Eq. (\ref{spectrum}).
Both the energy resolution function and the lower
energy threshold effects have been taken into account.
The spectra are normalized to have equal areas.
\label{pinched}}
\end{figure}

The spectra of neutrinos from the cooling stage are not
exactly thermal. Since the neutrino interaction cross-section
increases with energy, even for the same species of
neutrinos, the effective neutrinosphere radius increases with
increasing energies. Then, even if the neutrinos at their
respective neutrinospheres are in thermal equilibrium
with the matter, the spectrum gets
``pinched'', i.e. depleted at the higher as well as the lower
end of energies in comparison with thermal spectrum. (See
Fig.~\ref{pinched} for an illustration.)
It can be shown that pinching of the instantaneous spectrum is the
consequence of 
the following two facts: (i) the temperature inside the supernova
decreases with increasing radius and (ii) the density decreases
faster than $1/r$  \cite{lattimer}.
The pinching of the instantaneous spectrum can be extended
to the pinching of the time-integrated spectrum as long
as the time variation of the average energy of the spectrum
is small.
Numerical simulations of the neutrino spectra confirm
the pinching of spectra
\cite{schramm,janka-hill}.

One way to parametrize the pinched neutrino spectra is to
introduce an effective temperature $T_i$
and an effective degeneracy parameter $\eta_i$
(which has the same sign for neutrinos and antineutrinos 
and cannot be considered as the chemical potential)  
in the  Fermi-Dirac thermal spectrum for each species $i$: 
\beq
F_i^0(E) \propto \frac{E^2}{Exp(E/T_i - \eta_i) + 1} ~~.
\label{spectrum}
\eeq
For a pinched spectrum, $\eta_i > 0$.
The value of $\eta_i$ is the same for all $\nu_x$
species (neutrinos as well as antineutrinos, since they
have the same interactions),
and are in general different from
$\eta_e$ or $\eta_{\bar{e}}$. The value of $\eta_i$ need not
be constant throughout the cooling stage.
Typically, 
\beq
T_e \approx 3-4 \mbox{ MeV }~~,~~ T_{\bar{e}} \approx 5-6
\mbox{ MeV }~~,~~ T_x \approx 7-9 \mbox{ MeV }.
\eeq 
The typical values of $\eta_i$ are \cite{janka-hill,raffeltbook}
\beq
\eta_e \approx 3 - 5~~,~~ 
\eta_{\bar{e}} \approx 2.0 - 2.5~~,~~ 
\eta_x \approx 0 - 2~~.
\eeq
Notice that the strongest pinching is expected for the $\nu_e$
spectrum.
These values, however, are model-dependent and
we shall use them only as a guide.

In what follows, we shall calculate the fluxes of
electron neutrinos $F_e$,  electron antineutrinos, 
$F_{\bar{e}}$,  and the total flux of the non-electron
neutrinos, ``$4 F_x$'' at the Earth detectors.\\

%%%%%%%%%%%%%%%%%%%%%%%%%%%%%%%%%%%%%%%%%%%%%%%%%%%%%%%%%%%%%%%%%%%%
\subsection{Neutrino mass spectra}
\label{mass-spectra}
%%%%%%%%%%%%%%%%%%%%%%%%%%%%%%%%%%%%%%%%%%%%%%%%%%%%%%%%%%%%%%%%%%%

We consider the system of three active neutrinos
$\vec{\nu}_f \equiv (\nu_e, \nu_\mu, \nu_\tau)$ mixed in
vacuum, such that
\beq
\vec{\nu}_f = U \vec{\nu}~~,
\eeq
where $\vec{\nu} \equiv (\nu_1, \nu_2, \nu_3)$ is the
vector of  mass eigenstates
and 
\beq
U \equiv ||U_{fi}||  ~~, ~~ f = e,\mu,\tau, ~~,~~ i=1,2,3
\label{uei-def}
\eeq
is the mixing matrix.
We take
\beq
\Delta m_{32}^2 \equiv m_3^2 - m_2^2 = \Delta m^2_{atm}~~,
\eeq
so that the oscillations driven by $\Delta m^2_{23}$ solve
the atmospheric neutrino problem.
We identify
\beq
\Delta m_{21}^2 \equiv m_2^2 - m_1^2 = \Delta m^2_\odot~~,
\eeq
where $\Delta m^2_\odot$ is in one of the regions 
(SMA, LMA or VO) implied by the solar neutrino data 
(eqs.~\ref{sma-param},\ref{lma-param},\ref{vo-param}).

The key features of the spectra which play an important role in
the applications to the supernova neutrinos are:

1. the hierarchy of $\Delta m^2$:
\beq
|\Delta m^2_{32}| \approx |\Delta m^2_{31}| \gg |\Delta m^2_{21}|~~;
\eeq

2. the upper bound on $\Delta m^2$:
\beq
|\Delta m^2_{ij}| \lsim 10^{-2} \mbox{ eV}^2~~.
\eeq

Since  $\nu_\mu$ and $\nu_\tau$ are indistinguishable
in the supernova neutrino studies,   
the neutrino transitions are determined by the mixings of the
electron neutrino only, i.e. by the elements $U_{ei}$
(\ref{uei-def}) (see sect.~\ref{final-spec}  for more details). Moreover,
the three
$U_{ei}$'s are related by the unitarity condition
$\sum |U_{ei}|^2 = 1$, so that only two mixing elements are 
relevant, 
and one can use $|U_{e2}|$ and $|U_{e3}|$. 

The element $U_{e3}$ is small, as mentioned in the introduction. 
If $|U_{e3}|^2 \ll 1$, then $U_{e2}$ (and therefore, $U_{e1}$)
can be found from the solar neutrino data:
\beq
4 |U_{e2}|^2 |U_{e1}|^2  \approx 4 |U_{e2}|^2 (1 - |U_{e2}|^2) \approx
\sin^2 2\theta_\odot~~,
\eeq
where $\theta_\odot$ is the mixing angle determined in the
$2\nu$ analysis of the solar neutrino data.

The system 
is then determined by two pairs of parameters 
$(\Delta m^2_i, \sin^2 2\theta_i), ~i = L, H$,  where
\barr
(\Delta m^2_L, \sin^2 2\theta_L) & \equiapp &
(\Delta m^2_\odot, \sin^2 2\theta_\odot) \nonumber \\
(\Delta m^2_H, \sin^2 2\theta_H) & \equiapp &
(\Delta m^2_{atm}, 4 |U_{e3}|^2)~~.
\label{twopairs}
\earr     
Correspondingly, it can then be described by two points in the
$(\Delta m^2, \sin^2 2\theta)$--plane.

The current oscillation data do not determine the 
mass and flavor spectrum
completely. As we have already mentioned 
in the introduction,  the uncertainty is related to:

\begin{enumerate}

\item The discrete ambiguity in the solution of the solar
neutrino problem: The data favor three solutions
indicated in the introduction (SMA, LMA and VO), and
some other solutions ({\it e.g.} the low $\Delta m^2$ MSW solution
\cite{solar-fit,lowdmsq},
the trimaximal mixing solution \cite{trimax}), 
although disfavored, are not excluded.
The future solar neutrino experiments will remove this
ambiguity and sharpen the determination of the 
oscillation parameters.

\item The ambiguity in the sign of $\Delta m^2_{32}$
(and $\Delta m^2_{31}$): this determines the type of 
neutrino mass hierarchy.
We refer to the case 
\beq
\Delta m^2_{32} > 0 ~~,~~ \mbox{\it i.e.} ~~m_3 > m_2, m_1
\eeq
as to the spectrum with {\it normal} mass hierarchy and
to the case
\beq
\Delta m^2_{32} < 0 ~~,~~ \mbox{\it i.e.} ~~m_2, m_1 > m_3
\label{inv-h}
\eeq
as to the spectrum with {\it inverted} mass hierarchy.

The key difference between these two hierarchies is that,
in the normal hierarchy, the small $U_{e3}$-admixture of 
$\nu_e$ is in the heaviest state whereas in the inverted
hierarchy, this admixture is in the lightest state.

The type of hierarchy can in principle be established in 
future studies of atmospheric neutrinos. If the effects of
$\nu_e$ oscillations will be observed and the sign of the 
charged lepton produced by the atmospheric neutrinos
will be identified in the future experiments,
the studies of matter effects in the neutrino and
antineutrino channels will allow to establish the
sign of $\Delta m^2$. The sign can also be found from the
studies of matter effects in the future long baseline experiments,
in particular, with neutrino factories.

\item The value of $U_{e3}$ is unknown. In principle, future
atmospheric neutrino experiments and the long baseline
experiments will be able to measure or further restrict
$U_{e3}$.

\item In the case of the SMA and LMA solutions, 
$\Delta m^2_{21} > 0$. But the sign of $\Delta m^2_{21}$ is
undetermined in the VO solution. The feasibility of
resolving this ambiguity has been recently discussed
in \cite{kuo-last}.

\end{enumerate}

Summarizing, the ambiguity in the present analysis is
related to the solution of the solar neutrino problem, 
the type of hierarchy and the value of $U_{e3}$. The first
two ambiguities lead to six possible schemes of neutrino
masses and mixing. Within each scheme, the predictions depend on
the value of $U_{e3}$.

%%%%%%%%%%%%%%%%%%%%%%%%%%%%%%%%%%%%%%%%%%%%%%%%%%%%%%%%%%%%%%%%%%%%%%
\subsection{Neutrino conversion regions}
\label{trans}
%%%%%%%%%%%%%%%%%%%%%%%%%%%%%%%%%%%%%%%%%%%%%%%%%%%%%%%%%%%%%%%%%%%%%

In supernova, the transitions occur mainly in the resonance layers, 
where the density varies between
($\rho_{res} - \Delta \rho_{res}) \div 
(\rho_{res} + \Delta \rho_{res}$).
Here $\rho_{res}$ is the resonance density:
\beq
\rho_{res} \approx \frac{1}{2 \sqrt{2} G_F} 
\frac{\Delta m^2}{E} 
\frac{m_N}{Y_e} \cos 2\theta ~~,
\label{rho-r}
\eeq
$G_F$ is the Fermi constant,
$m_N$ is the mass of the nucleon,
$E$ is the neutrino energy and $Y_e$
is the electron fraction -- the number of electrons
per nucleon\footnote{It is worthwhile to 
introduce the resonance density for
``VO'' parameters  too,  although the dominating effect
could be the vacuum oscillations. We shall consider
the limit of $\sin^2 2\theta  \to 1$ as a special case.}. 
For small vacuum mixing, 
the width of the resonance layer equals
\beq
2 \Delta \rho_{res} \approx 2 \rho_{res} \tan 2\theta~~.
\label{res-width}
\eeq
Using (\ref{rho-r}),
the resonance matter density can be written as
\beq
\rho_{res} \sim 1.4 \times 10^{6} \mbox{ g/cc}
\left( \frac{\Delta m^2}{1 \mbox{ eV}^2} \right)
\left( \frac{10 \mbox{ MeV}}{E} \right)
\left( \frac{0.5}{Y_e} \right)
\cos 2\theta~~.
\label{rho-res}
\eeq

There are two resonance layers.  
The layer at higher densities ($H$-resonance layer), which corresponds to
$\Delta m^2_{atm}$, is at
\beq
\rho_H \sim 10^3 - 10^4 \mbox{ g/cc}~~,
\label{rhoh}
\eeq
and the layer at lower densities ($L$-resonance layer),
characterised by $\Delta m^2_\odot$, is at 
\beq
\rho_L = \left\{ 
\begin{array}{rcl}
5 - 15 & \mbox{ g/cc } & \mbox{ for SMA} \\
10 - 30 & \mbox{ g/cc } & \mbox{ for LMA} \\
< 10^{-4} & \mbox{ g/cc } & \mbox{ for VO} ~~.
\end{array}
\right.
\label{rhol}
\eeq

The regions where the neutrino transitions occur are thus
far outside the core of the star -- in the outer layers 
of the mantle. Therefore

\begin{itemize}

\item 
the transitions do not influence
the dynamics of collapse or the cooling of the core;

\item 
the r-processes, which occur at 
$\rho \gsim 10^5 - 10^6$ g/cc, are also not
affected; 

\item 
the shock wave does not influence the neutrino conversions
(indeed, during the time of cooling by neutrino emission 
($t \sim 10$  sec),
the shock wave can only reach layers with densities 
$\rho \gsim 10^6$ g/cc \cite{mayle,colgate}); 

\item
the density profile encountered by the neutrinos during their
resonant conversions is
almost  static, and the same as that of the
progenitor star. 

\end{itemize}

In the region with densities $\gsim 1$ g/cc,
the electron fraction is almost constant
and the density profile can be approximated by
\cite{notzold, janka-hill, bbb}
\beq
\rho Y_e \approx 2 \cdot 10^{4} \mbox{ g/cc} 
\left( \frac{r}{10^9 \mbox{ cm }} \right)^{-3}~~,~~
\mbox{ for } \rho \gsim 1 \mbox{ g/cc }.
\label{profile}
\eeq 
For $\rho \lsim 1$ g/cc,  
the fraction of hydrogen increases and $Y_e$ becomes larger than 0.5. 
The exact shape of the density profile depends on the details of
the composition of the star.

%%%%%%%%%%%%%%%%%%%%%%%%%%%%%%%%%%%%%%%%%%%%%%%%%%%%%%%%%%%%%%%%%%%%%
\subsection{Factorization of dynamics}
\label{fact-dyn}
%%%%%%%%%%%%%%%%%%%%%%%%%%%%%%%%%%%%%%%%%%%%%%%%%%%%%%%%%%%%%%%%%%%%%

The hierarchy of $\Delta m^2$, and therefore the hierarchy
of the densities of the resonance layers, leads to the
``factorization'' of dynamics of conversion:  
the transitions in the two resonance layers can be 
considered independently and each transition is reduced to
a two neutrino problem. 
Indeed, in the $H$-resonance region, the mixing $U_{e2}^m$
associated with $\Delta m^2_\odot$ is suppressed by matter.
The suppression factor is
\beq
\frac{U_{e2}^m}{U_{e2}} \sim \frac{\rho_L}{\rho_H} 
\lsim 10^{-2}~~.
\eeq
Correspondingly, the effects driven by $\Delta m^2_\odot$
are suppressed by more than two orders of magnitude.

In the $L$-resonance region, the mixing associated with
$\Delta m^2_{atm}$ coincides with the vacuum mixing:
$U_{e3}^m \approx U_{e3}$. The matter corrections are
strongly suppressed:
\beq
U_{e3}^m = U_{e3} (1 + {\cal O}(\xi))~~,~~
~~~~~ \xi \approx \frac{\rho_L}{\rho_H} 
\lsim 10^{-2}~~.
\eeq
That is, the mixing associated with $\Delta m^2_{atm}$ 
is almost constant, and therefore the level
$\nu_3$ practically does not participate in the
dynamics. By an appropriate redefinition of
the fields, the problem can be reduced to a two
state problem. The state $\nu_3$ decouples from the
rest of the system, producing just
an averaged oscillation effect \cite{kuo-rmp}.

If the mixing $U_{e3}$ is very small, the resulting
survival probability of $\nu_e$ ($\bar{\nu}_e$)
is also factorized \cite{kuo-prd37,ms-ppnp}:
\beq
[P(\nu_e \to \nu_e)]_{total} = [P(\nu_e \to \nu_e)]_H 
~\times~[P(\nu_e \to \nu_e)]_L ~ + {\cal O}(|U_{e3}|^2)~~, 
\label{frac}
\eeq
and similarly for $\bar{\nu}_e$.

%%%%%%%%%%%%%%%%%%%%%%%%%%%%%%%%%%%%%%%%%%%%%%%%%%%%%%%%%%%%%%%%%
\subsection{Adiabatic, non-adiabatic and transition regions}
\label{adia}
%%%%%%%%%%%%%%%%%%%%%%%%%%%%%%%%%%%%%%%%%%%%%%%%%%%%%%%%%%%%%%%%%

The dynamics of transitions in each resonance layer is 
determined by the adiabaticity parameter $\gamma$
\cite{l-z, ms}:
\beq
\gamma \equiv \frac{\Delta m^2}{2 E} 
\frac{\sin^2 2\theta}{\cos 2\theta}
\frac{1}{(1/n_e)(dn_e/dr)}~~,
\label{gamma-def}
\eeq
such that the ``flip'' probability -- the probability
that a neutrino in one matter eigenstate jumps
to the other matter eigenstate -- is
\beq
P_f = exp \left(-\frac{\pi}{2} \gamma \right)~~
\label{pf-def}
\eeq
as given by the Landau-Zener formula \cite{l-z}.
Adiabatic conversion corresponds to $\gamma \gg 1$,
{\it i.e.} to a very small flip 
probability~\footnote{The Landau-Zener formula is valid for a linear
variation of
density in the resonance region and a small mixing angle 
$\theta$. For an arbitrary density distribution and
mixing angle, the Landau-Zener formula (\ref{pf-def}) 
gets modified to
\cite{kuo-prd}:
\beq
P_f = \frac{ exp(-\frac{\pi}{2} \gamma F) -
                exp(-\frac{\pi}{2} \gamma F/\sin^2 \theta)}
        {1 - exp(-\frac{\pi}{2} \gamma F/\sin^2 \theta)}~~,
\label{exact-pf}
\eeq
where $F$ is a function of the density profile and the mixing
angle. The adiabaticity condition is, approximately, 
$\gamma F \gg 1$.
At small mixing angles and linear density variation in the
resonance region,
$F \approx 1$ and (\ref{exact-pf}) reduces to (\ref{pf-def}).}.

Let us consider the density profile of the form
\beq
\rho = A/r^n~~,
\label{rho-rn}
\eeq
where A is the proportionality constant. 
From (\ref{gamma-def}), we get the 
adiabaticity parameter for this profile as
\beq
\gamma = \frac{1}{2n} 
\left( \frac{\Delta m^2}{E} \right) ^{1 - 1/n}
\frac{\sin^2 2 \theta}{(\cos 2 \theta)^{1+1/n}}
\left( \frac{2 \sqrt{2} G_F  Y_e}{m_N} A \right)^{1/n}~~.
\label{gamma-dep}
\eeq
Here we have used the resonance condition to express
$r$ through the oscillation parameters.
Note that the dependence of $\gamma$ on the absolute scale of 
density, $A$, is rather weak: 
\beq
\gamma \propto A^{1/n}~~.
\label{a1/n}
\eeq
For $n=3$, the change of $A$ by one order of magnitude leads to
the change of $\gamma$ by a factor of 2.

For a fixed density scale $A$, the value of $\gamma$ depends on
the power index $n$ as
\beq
\gamma \propto 1/n~~,
\label{1/n}
\eeq
so that a variation  of $n$ between 2 to 4 leads to the variation
in $\gamma$ by a factor of 2. 

In turn, the uncertainty of
a factor of 2 in $\gamma$ is equivalent to 
the change
of $\sin^2 2 \theta$ by a factor of 2
(at small values of $\theta$), or the change
of $\Delta m^2$ by a factor of $2^{n/(n-1)} \sim 3 - 4$.

The lines of equal $\gamma$ (and therefore, equal $P_f$) on
the ($\Delta m^2 - \sin^2 2 \theta$) -- plot are determined by
\beq
(\Delta m^2)^{1-1/n} \sin^2 2 \theta = \mbox{\it const }
\eeq
for small $\theta$. 

For $n=3$ 
which will be used in the further calculations, we get
\beq
\Delta m^2 = \mbox{\it  const }/\sin^3 2\theta~~.
\eeq 
From (\ref{pf-def}) and (\ref{gamma-dep}), the flip probability
as a function of energy can be written as \cite{buccella}
\beq
P_f = \mbox{ Exp} \left[ - \left( \frac{E_{na}}{E} \right)^{2/3}
\right]~~,
\eeq
where
\beq
E_{na} = \left( \frac{\pi}{12} \right) ^{3/2}
\frac{\Delta m^2 \sin^3 2\theta}{\cos^2 2\theta}
\left( \frac{2 \sqrt{2} G_F Y_e}{m_N} A \right)^{1/2}~~.
\label{ena-def}
\eeq
\begin{figure}[htb]
\hbox to \hsize{\hfil\epsfxsize=10cm\epsfbox{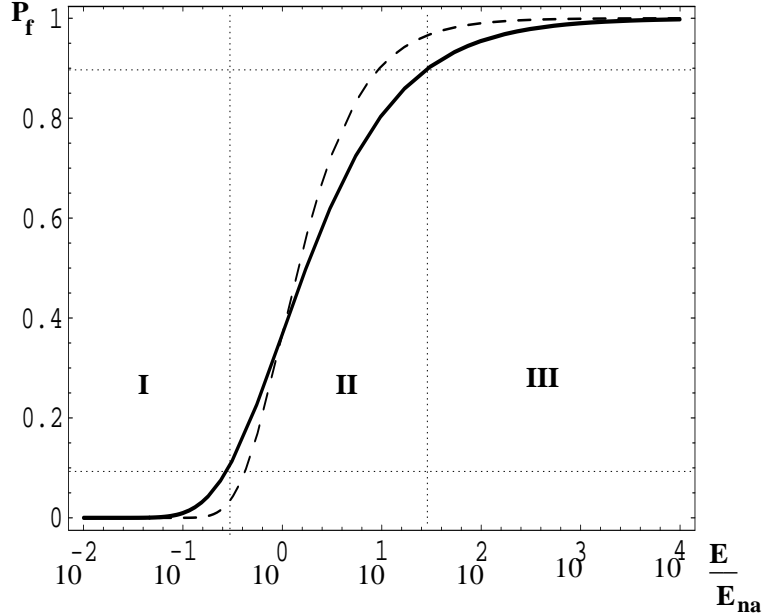}\hfil}
\caption{~~The energy dependence of $P_f$
on $E/E_{na}$. The solid line is for
the density profile $\rho \sim r^{-3}$, whereas the
dashed line is for
the density profile $\rho \sim e^{-r}$. 
\label{pfvar}}
\end{figure}
The dependence of $P_f$ on $E/E_{na}$ is shown in 
Fig.~\ref{pfvar}. One can divide the whole  range of energy 
in three parts: 

For $E/E_{na} < 10^{-1}$ (region I), 
we get $P_f \approx 0$. In this range,
pure adiabatic conversion occurs. 

For $E/E_{na} > 10^2$ (region III), the flip probability is close to
1, which corresponds to a strong violation of adiabaticity.

In the transition region 
$E/E_{na} = 10^{-1} - 10^2$ (region II),
$P_f$ increases with the neutrino energy. 
This region spans almost
three orders of magnitude in energy, 
which is substantially larger than
the range of energies in the neutrino spectrum.
Notice that for the exponential density distribution
(which is the case inside the sun), the transition region is
narrower (about two orders of magnitude) and 
correspondingly, the energy dependence in this
region is stronger.

\begin{figure}[htb]
\hbox to \hsize{\hfil\epsfxsize=14cm\epsfbox{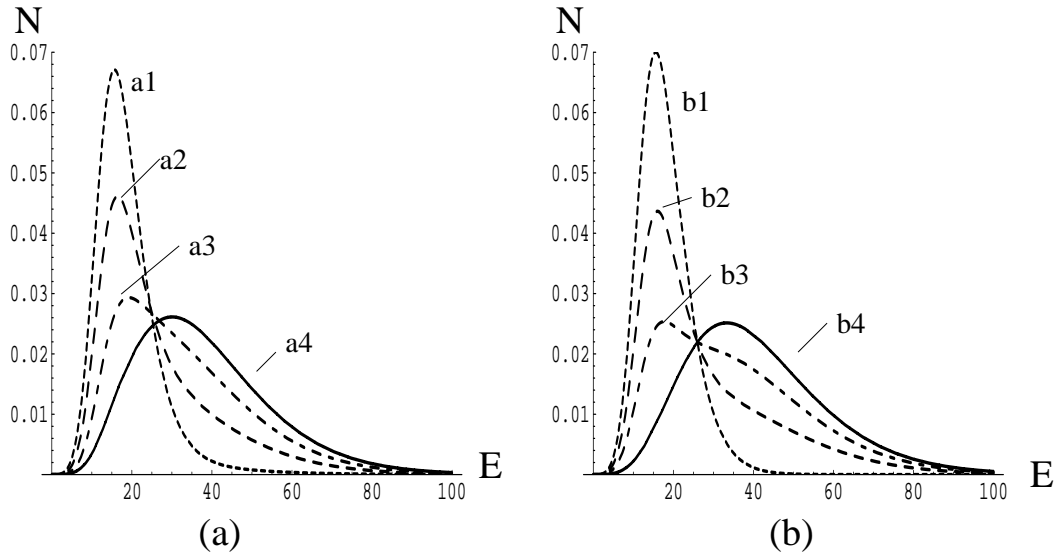}\hfil}
\caption{~~The number of $\nu_e-N$ charged current events,
taking into account the energy dependence of the flip probability.
The parameters for the original spectra are taken to be
$T_e = 3$ MeV, $\eta_e = 3$, $T_x = 8$ MeV, $\eta_x = 1$. 
In (a), the flip probability is $P_f = Exp[-(E_{na}/ E)^{2/3}]$ with
(a1): $E_{na}=0.05$, (a2): $E_{na}=2$, (a3): $E_{na}=10$, 
(a4): $E_{na}=50$ MeV.
In (b), the effective flip probability $\langle P_f \rangle$ is
(b1): 1.0, (b2): 0.85, (b3): 0.6, (b4): 0.0
\label{e-dep}}
\end{figure}

The observable part of the 
supernova neutrino spectrum lies mainly between the 
energies of 5 and 50 MeV, {\it i.e.} it spans about one order of 
magnitude. If the spectrum is in region I, completely
adiabatic conversion occurs for the whole spectrum.
In the region II, the conversion depends on energy, however
the dependence is not strong over the relevant
range of energies. 
The average energies of any two neutrino species differ 
by a factor of less than 3, and for $\Delta E / E \sim 3$,
the variation in the flip probability is $\Delta P_f \leq
0.2$ (from Fig.~\ref{pfvar}).
In the first approximation, the final spectrum can then be
characterized by an {\it average} or {\it effective} flip probability.
This is illustrated in Fig.~\ref{e-dep}:
the spectra of the number of events taking into
account the
energy dependence of $P_f$ are shown in (a) and  the
spectra with an effective flip probability $\langle P_f \rangle$
are shown in (b). It may be observed that the spectra
with an appropriate value of $\langle P_f \rangle$
can mimic the features of the actual spectra.

\begin{figure}[htb]
\hbox to \hsize{\hfil\epsfxsize=10cm\epsfbox{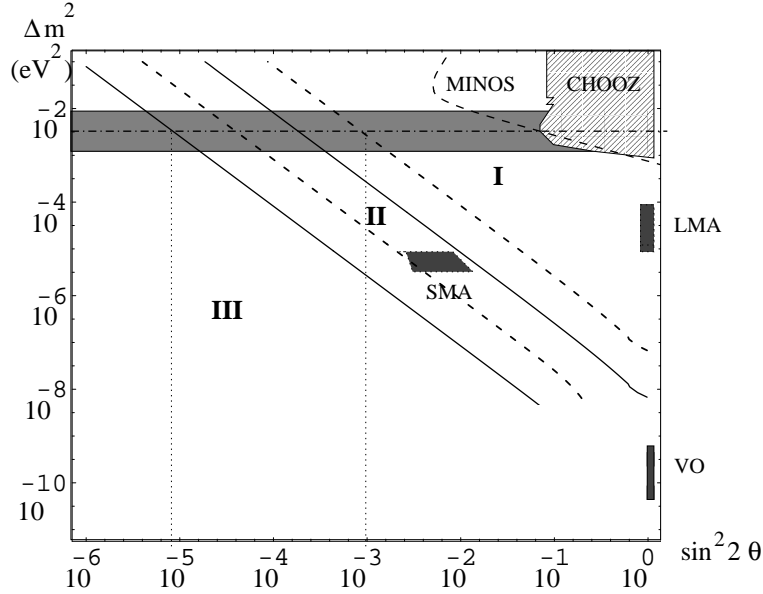}\hfil}
\caption{~~The contours of equal
flip probability $P_f$.
The solid lines denote the contours of flip probability
for a 5 MeV neutrino: the line on the left stands for 
$P_f = 0.9$ (highly non-adiabatic transition) 
and the line on the right stands for
$P_f = 0.1$ (adiabatic transition).
The dashed lines represent the corresponding flip probabilities
for neutrinos with energy 50 MeV.
SMA, LMA and VO
correspond to the solutions of the solar
neutrino anomaly. The two vertical lines
indicate the values of $4 |U_{e3}|^2 = \sin^2 2\theta$ 
lying on the
borders of the adiabatic, non-adiabatic and
transition regions for $\Delta m^2$ corresponding to
the best fit value of the atmospheric neutrino solution. 
\label{adia-box}}
\end{figure}

In Fig.~\ref{adia-box}, we show the contours of equal
flip probability $P_f$ (\ref{exact-pf})
in the ($\Delta m^2 - \sin^2 2\theta$) plot
for two different energies on the
borders of the observable spectrum.
We also show the parameter ranges 
which explain the
solar and atmospheric neutrino data.
The dark band (the ``atmospheric neutrino band'')
corresponds to the
allowed range of $\Delta m^2_{31}$.
The rightmost
part of this band 
is excluded by the CHOOZ experiment
\cite{chooz}. The range of $|U_{e3}|^2$ that can be
probed by the long baseline experiment MINOS \cite{minos}
is also shown.

The contours of $P_f = 0.1$ and $P_f = 0.9$ divide the plot
into three regions, corresponding to the three
regions in fig.~\ref{pfvar}: 

\begin{itemize}

\item 
The ``adiabatic region'' (I) is the region above the contour with
$P_f = 0.1$, where  adiabaticity
is well satisfied and strong flavor conversions
occur.

\item 
The ``transition region'' (II) is the region between
$P_f=0.1$ and $P_f = 0.9$ contours. Here the adiabaticity
is partially broken and the transitions are not complete.
Moreover, the extent of transitions depends on the energy.

\item 
The ``non-adiabatic'' region (III)  lies below
the $P_f = 0.9$ contour.
The neutrino conversions are practically absent.

\end{itemize}

Since the adiabaticity breaking increases with $E$ and 
decreases with the increase of $\Delta m^2$ and $\sin^2 2\theta$,
the lines of equal $P_f$ for $E=50$ MeV are shifted to
larger $\Delta m^2$ and  $\sin^2 2\theta$ relative to the
lines for $E=5$ MeV. 
The dependence of the contours of equal $P_f$ on the density
profile is rather weak, as can be seen from the previous discussion. 
Even with conservative estimates, the borders of the regions have an 
uncertainty of a factor of 2 in $\sin^2 2\theta$ for
a given value of $\Delta m^2$.

As follows from Fig.~\ref{adia-box}, the LMA solution
lies in the adiabatic region I\footnote{The LOW solution also
lies in the adiabatic region, so all the results for the
LMA scenario are also valid for the LOW scenario.}. 
The SMA solution is in the 
transition region. The VO solutions are either in the
transition region or in the non-adiabatic region, which
depends essentially on the density profile 
in the outermost layers of the star
($\rho \lsim 1$ g/cc) and the precise value of 
$\Delta m^2$.

As was described in sect. \ref{mass-spectra}, 
each neutrino mass and flavor spectrum can be 
represented by two points (\ref{twopairs}) in the 
$(\Delta m^2, \sin^2 2\theta)$ plot (Fig.~\ref{adia-box}).
One point, corresponding to 
($\Delta m^2_{31}, \sin^2 2\theta_{e3}$),
should lie in the atmospheric neutrino band, 
and the other point corresponding to
($\Delta m^2_{21}, \sin^2 2\theta_{e2}$)
should lie in one of  the  ``islands''
corresponding to the solutions of the solar neutrino
problem.  
These two points characterize the layers $H$ and $L$
(\ref{rhoh},\ref{rhol}) respectively.
Let $P_H (\bar{P}_H)$ and $P_L (\bar{P}_L)$
be the probabilities that the neutrinos (antineutrinos)
jump to another matter eigenstate in these two layers.
The extent of conversion is determined by the values of
these four flip probabilities.

From Fig.~\ref{adia-box}, we conclude that 
the $H$-resonance is in the adiabatic range (region I)
for
\beq
\sin^2 2\theta_{e3} = 4 |U_{e3}|^2 \gsim 10^{-3}~~,
\label{reg1limit}
\eeq
and in the transition region (region II) for
\beq
\sin^2 2\theta_{e3} \sim 10^{-5} - 10^{-3}~~.
\label{reg2limit}
\eeq
As we shall see in sect. \ref{normal} and ~\ref{inverted},
the features of the final spectra depend strongly on 
the region in which 
the $H$ resonance lies. 
According to  (\ref{reg2limit}), the supernova neutrino spectra
are sensitive to as low values of $4 |U_{e3}|^2$ as 
$10^{-3} - 10^{-5}$. 
This is more than two orders of magnitude better
than the current bounds \cite{chooz} or those expected from the 
planned long baseline experiments.

%%%%%%%%%%%%%%%%%%%%%%%%%%%%%%%%%%%%%%%%%%%%%%%%%%%%%%%%%%%%%%%%%%%%%%
\subsection{The level crossing schemes and initial conditions}
\label{levelx}
%%%%%%%%%%%%%%%%%%%%%%%%%%%%%%%%%%%%%%%%%%%%%%%%%%%%%%%%%%%%%%%%%%%%%%

In the basis of flavor eigenstates
$(\nu_e, \nu_\mu,\nu_\tau)$, 
the evolution of neutrinos 
at  densities $\rho < 10^{6}$ g/cc 
relevant for neutrino conversion (see below)
is described by
a Schr\"odinger-like equation with 
the effective Hamiltonian 
\barr
H & = & \frac{{\cal M}^2}{2E} + {\cal V} \nonumber \\
& = & \frac{1}{2 E} \left( 
\begin{array}{lcc} 
m_{ee}^2 + 2 E V & m_{e\mu}^2 & m_{e\tau}^2  \\
m_{e\mu}^2 & m_{\mu\mu}^2 & m_{\mu\tau}^2  \\
m_{e\tau}^2 & m_{\mu\tau}^2 & m_{\tau\tau}^2 
\end{array}
\right)~~,
\label{hamilt}
\earr
where ${\cal V} \approx Diag(V, 0, 0)$, 
and $V =\sqrt{2} G_F n_e$ is the effective potential 
for the electron neutrinos due to their charged current
interactions with electrons.

Since any rotation in the ($\nu_\mu - \nu_\tau$)
subspace does not affect the physics,
it is convenient to perform a rotation of the
neutrino states $(\nu_e, \nu_\mu,\nu_\tau) \to
(\nu_e, \nu_{\mu'},\nu_{\tau'})$, which diagonalizes the
$(\nu_\mu,\nu_\tau)$ submatrix of (\ref{hamilt}) \cite{kuo-prl}.
(The potential $V$ appears only in the element $H_{ee}$, 
and hence is not affected by this rotation.) 
The effective Hamiltonian in the 
new basis becomes
\beq
H = \frac{1}{2E}\left( 
\begin{array}{lcc} 
m_{ee}^2 + 2 EV & m_{e \mu'}^2 & m_{e\tau'}^2  \\
m_{e\mu'}^2 & m_{\mu' \mu'}^2 & 0  \\
m_{e\tau'}^2 & 0 & m_{\tau' \tau'}^2 
\end{array}
\right)~~.
\label{hamilt-new}
\eeq  

At  $V \gg m^2_{ij}/(2 E)$, 
the off-diagonal terms can be neglected and 
the Hamiltonian (\ref{hamilt-new}) becomes diagonal:
\beq
H \approx {\rm Diag} (V, m^2_{\mu' \mu'}, m^2_{\tau' \tau'})~~.
\eeq
That is, the basis states $(\nu_e, \nu_{\mu'}, \nu_{\tau'})$
are the matter eigenstates.
These are the states that arrive at the conversion regions 
as independent (incoherent) states and transform in this
region independently.

Notice that a difference between the  potentials of 
$\nu_{\mu}$ and $\nu_{\tau}$
appears in the second order in the weak interactions  
due to difference of masses of the $\mu$ and $\tau$ 
charged leptons \cite{botella}: 
\beq
V_{\mu \tau} \approx V 
\frac{3 G_F m_{\tau}^2}{2\sqrt{2}\pi^2 Y_e}
\left(\ln \frac{m_W^2}{m_{\tau}^2} - 1 + \frac{Y_n}{3}\right) 
\approx 10^{-4} ~V~, 
\eeq
where 
$m_{\tau}$ is the $\tau$ mass, 
$m_W$ is the $W$-boson mass,  $Y_e$ and  $Y_n$
are the numbers of electrons and the neutrons per nucleon 
respectively.  Therefore a complete form of the matrix of 
potentials is ${\cal V} \approx Diag(V, 0, V_{\mu \tau})$. 

The potential $V_{\mu \tau}$ becomes important at high densities. 
One has 
$$
V_{\mu \tau} \sim \Delta m^2_{atm}/2E \approx  2 m_{\mu \tau}^2/2E  
\mbox{ at } \rho_{\mu \tau} \sim 10^{7} - 10^{8} \mbox{ g/cc}.
$$ 
At $\rho \gg \rho_{\mu \tau}$, and in particular in the 
region of the neutrinosphere,  
the potentials $V$ and $V_{\mu \tau}$ dominate over the
other terms in the Hamiltonian, and the Hamiltonian 
becomes approximately diagonal:
$H \approx   {\cal V} \approx {\rm Diag}(V, 0, V_{\mu \tau})$. 
This means that at high densities the flavor states 
coincide with the eigenstates in medium.

Let us 
recall that the non-electron neutrinos are produced in the 
neutral current processes which are flavor blind,
{\it i.e.}  they produce
a coherent mixture of matter eigenstates.
This coherence, however, disappears in the evolution
that follows.
The $\nu_e$ departs from the coherent state due to the large
potential $V$, whereas the coherence of
$\nu_{\mu}$ and $\nu_{\tau}$ is broken by  $V_{\mu \tau}$.

In the interval  of densities where 
$V_{\mu \tau} \ll  \Delta m^2_{atm}/2E \ll V$, the potential 
$V_{\mu \tau}$ can be neglected, so that we arrive at the Hamiltonian 
(\ref{hamilt-new})
with eigenstastes  
$(\nu_e, \nu_{\mu'}, \nu_{\tau'})$. In the region 
$V_{\mu \tau} \sim   \Delta m^2_{atm}/2E$, the level crossing occurs
which leads to the transitions $\nu_{\mu} \rightarrow \nu_{\tau}'$ 
and  
$\nu_{\tau} \rightarrow \nu_{\mu}'$. 
In the antineutrino channel, we have 
$\bar{\nu}_{\mu} \rightarrow \bar {\nu}_{\mu}'$
and $\bar{\nu}_{\tau} \rightarrow \bar {\nu}_{\tau}'$.

Since the initial fluxes of $\nu_{\mu}$ and $\nu_{\tau}$  are equal
($F_{x}^0$), we get that  
the fluxes of  $\nu_{\mu'}$ and $\nu_{\tau'}$
will be also equal ($F_{x}^0$). Therefore  
($\nu_e, \nu_{\mu'},\nu_{\tau'}$) with fluxes 
$(F_{e}^0, F_{x}^0, F_{x}^0)$ and correspondingly 
($\bar{\nu}_e, \bar{\nu}_{\mu'}, \bar{\nu}_{\tau'}$) 
with fluxes $(F_{\bar e}^0, F_{x}^0, F_{x}^0)$ can 
be considered as the initial state in our task. \\

\begin{figure}
\hbox to \hsize{\hfil\epsfxsize=16 cm\epsfbox{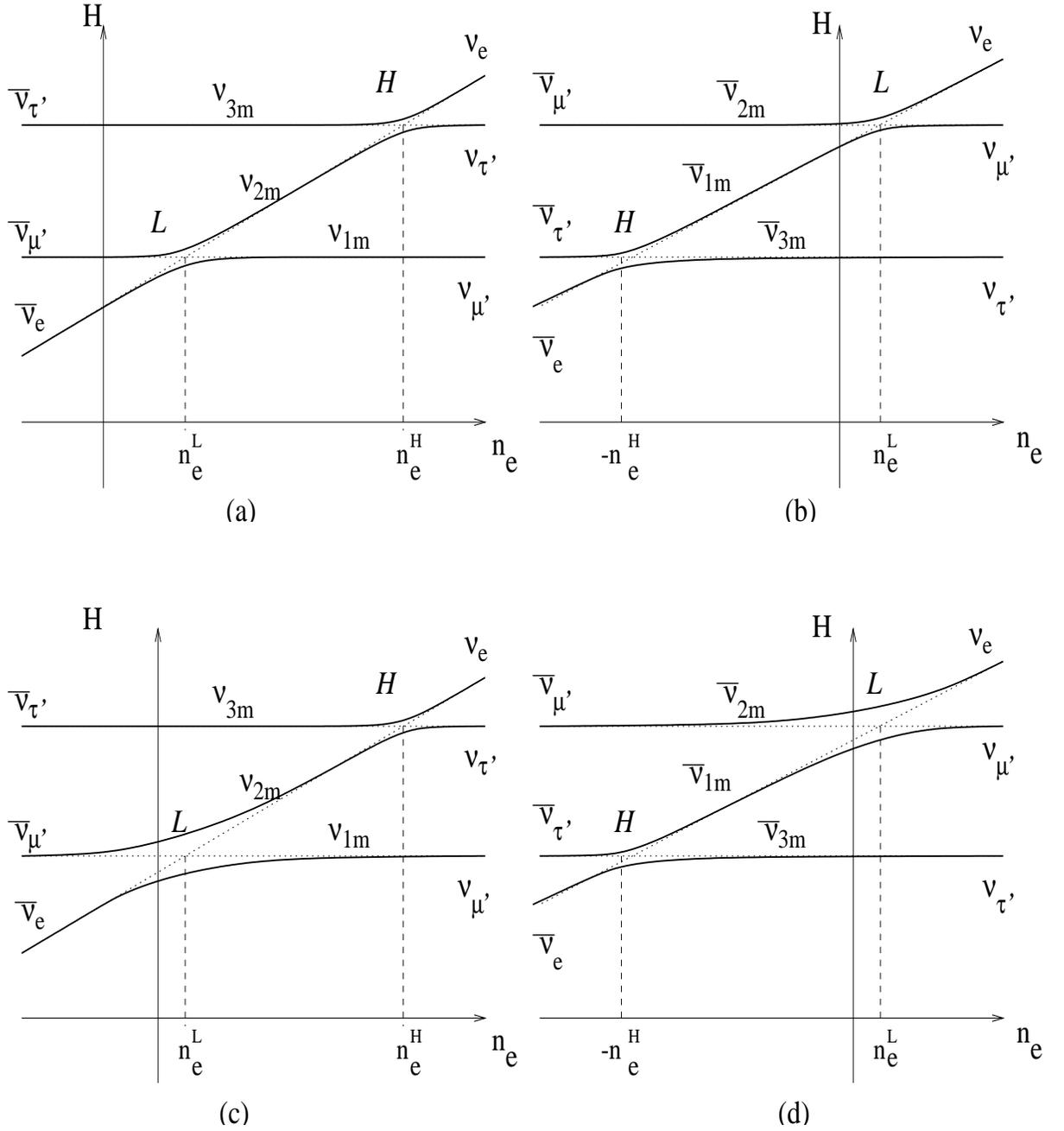}\hfil}
\caption{~~The level crossing diagrams for 
(a) the normal mass hierarchy and small $\theta_\odot$, 
(b) the inverted mass hierarchy and small $\theta_\odot$,
(c) the normal mass hierarchy and large $\theta_\odot$, 
(d) the inverted mass hierarchy and large $\theta_\odot$.
Solid lines show the eigenvalues of the effective
Hamiltonian as functions of the electron number
density. The dashed lines correspond to energies of
flavor levels $\nu_e$, $\nu_{\mu'}$, and  $\nu_{\tau'}$.
The part of the plot with $n_e < 0$  corresponds to the
antineutrino channel.
\label{generic}}
\end{figure}

The Hamiltonian (\ref{hamilt-new}) 
allows us easily to construct the level crossing scheme.
In Fig.~\ref{generic}, we show the generic level crossing
diagrams for the normal and inverted mass hierarchies,
for small $\theta_\odot$ (SMA) as well as large $\theta_\odot$
(LMA or VO). 
The diagonal elements of the Hamiltonian (\ref{hamilt-new}),
$H_{ii}(n_e)~~(i = e,\mu',\tau')$, determine the energies
of the flavor states shown by the dotted lines.
The crossing of these levels indicates a resonance.
$H$ and $L$ are the two resonances at higher and lower
densities respectively (\ref{rhoh},\ref{rhol}).
The solid lines represent the
eigenvalues of the Hamiltonian (\ref{hamilt-new}).

In the case of antineutrinos,
the effective potential $V$ for the  $\bar{\nu}_e$
has the opposite sign : $V = -\sqrt{2} G_F n_e$.
The antineutrinos can then be represented on the
same level crossing diagram, as neutrinos
travelling through matter with
``effectively'' negative $n_e$. The half-plane with
positive values of $n_e$ then describes neutrinos and
the half-plane
with negative values of $n_e$ describes antineutrinos.

The neutrinos (antineutrinos) are produced inside
a supernova in regions of high matter density.
On their way towards the earth,
they travel through a medium with {\it almost}
monotonically decreasing density, towards the vacuum
where both neutrinos and antineutrinos
have vanishing effective potentials. This corresponds
to starting at the right (left) extreme ends of
the $n_e$ axis in Fig.~\ref{generic}, and
moving towards $n_e=0$.

The $H$ resonance lies in the neutrino channel for the normal
hierarchy and in the antineutrino channel for the inverted 
hierarchy. The $L$ resonance lies in the neutrino channel for
both the hierarchies as long as the solar neutrino solution is
SMA or LMA. For the VO solution, the $L$ resonance may lie
in either of the two channels, neutrinos or antineutrinos.

%%%%%%%%%%%%%%%%%%%%%%%%%%%%%%%%%%%%%%%%%%%%%%%%%%%%%%%%%%%%%%%%%%%%%
\section{Conversion probabilities and the neutrino fluxes at the
detectors. The case of normal mass hierarchy}
\label{final-spec}
%%%%%%%%%%%%%%%%%%%%%%%%%%%%%%%%%%%%%%%%%%%%%%%%%%%%%%%%%%%%%%%%%%%%%%

In this section, we derive general expressions for the
transition probabilities using the
level crossing scheme for the normal mass hierarchy as 
shown in Figs.~\ref{generic}a and \ref{generic}c. (The inverted hierarchy will
be discussed separately in Sec.~\ref{inverted}.)

%%%%%%%%%%%%%%%%%%%%%%%%%%%%%%%%%%%%%%%%%%%%%%%%%%%%%%%%%%%%%%%%%%
\subsection{Probabilities of conversion inside the star}
%%%%%%%%%%%%%%%%%%%%%%%%%%%%%%%%%%%%%%%%%%%%%%%%%%%%%%%%%%%%%%%%%%

As has been discussed in sec.~\ref{levelx}, the neutrino
fluxes arise from the central part of the star, in the
region of high density. For $\rho \gg \rho_H, \rho_L$,
where all the mixings are highly suppressed,
the flavor states ($\nu_e, \nu_{\mu'}, \nu_{\tau'}$)
coincide with the eigenstates in the medium :
\beq
\nu_{3m}=\nu_e ~~,~~ 
\nu_{2m}=\nu_{\tau'} ~~,~~ 
\nu_{1m}=\nu_{\mu'}~~.
\label{largerho}
\eeq
The original fluxes of neutrino eigenstates in 
the medium 
equal
\beq
F_{1m}^0 =   F_x^0 ~~,~~
F_{2m}^0  =   F_x^0 ~~,~~
F_{3m}^0  =  F_e^0~~.
\eeq

Let us calculate the fluxes of the mass eigenstates 
$\nu_i$ at the surface of the star. These 
states, being the eigenstates of the Hamiltonian in vacuum,
travel independently to the surface of the earth.

Taking into account that the dynamics of 
transitions in the  two
resonance layers are independent (see Sec.~\ref{fact-dyn}), 
the fluxes of neutrino mass eigenstates at the
surface of the star can be written down directly
by tracing the path of the neutrinos in the level
crossing diagram. 
We find the modifications of fluxes in terms of the
flip probabilities $P_H, P_L, \bar{P}_H$ and $\bar{P}_L$
introduced in Sec.~\ref{adia}.

Let us first calculate the flux of $\nu_1$
at the surface of the star. There are
three independent contributions to this flux, from the initial
$\nu_e, \nu_{\mu'}$ and $\nu_{\tau'}$ fluxes.
The probability that the original state $\nu_e$ (which
coincides with $\nu_{3m}$ in the production region)
arrives at the surface of the star as $\nu_1$ is
$P_H P_L$, since the state has to flip to the other
matter eigenstate at both the resonances.
The contribution to the final $\nu_1$ flux from
the original $\nu_e$ flux is then $P_H P_L F_e^0$.
Similarly, the contributions from
the original $\nu_{\mu'}$ and $\nu_{\tau'}$ equal
$(1-P_L) F_x^0$ and $P_L (1 - P_H) F_x^0$ respectively.
The total $\nu_1$ flux at the surface of the star 
equals the sum of the three contributions:
\beq
F_1  =  P_{H} P_{L} F_e^0 + (1 - P_{H} P_{L}) F_x^0~~.
\label{f1star}
\eeq
Similarly, the fluxes of neutrino mass eigenstates
$\nu_2$ and $\nu_3$ arriving at the surface of the
star are
\barr
F_2 & = & (P_{H} - P_{H} P_{L}) F_e^0 +
        ( 1 -P_{H} + P_{H} P_{L}) F_x^0 ~~,\nonumber \\
F_3 & = & (1 - P_{H}) F_e^0 + P_{H} F_x^0 ~~.
\label{fstar}
\earr

Due to the divergence of the wavepackets, 
any coherence between the mass eigenstates is
lost on the way to the earth.
Indeed,
over a distance L, the two wavepackets corresponding to two mass
eigenstates with a given $\Delta m^2$ and having an energy $E$ 
separate from each other by a distance
\beq
\Delta L =  \frac{\Delta m^2}{2 E^2}~L~~.
\eeq
Even for the smallest $\Delta m^2 \sim 10^{-10}$ eV$^2$,
for $E \sim 10$ MeV and $L \sim 10$ kpc $\sim 10^{22}$ cm, 
we get $\Delta L \sim 10^{-2}$ cm. The lengths of the 
individual wavepackets are $\sigma \lsim 1/T \sim 10^{-11}$
cm (where $T$ is the temperarture of the production region) is much
smaller. 

The spread of the wavepackets  implies that 
the neutrinos arrive at the surface of the earth as
incoherent  fluxes of the mass eigenstates. Upto a
geometrical factor of $1/L^2$, they coincide with the fluxes
given in (\ref{f1star},\ref{fstar}). 
We can rewrite them as
\beq
F_i = a_i F_e^0 + (1 - a_i) F_x^0~~,
\label{fi}
\eeq
with
\beq
a_1 = P_H P_L, \ a_2 = P_H (1 - P_L), \ a_3 = 1 - P_H~~.
\label{ai-def}
\eeq
The factor of $1/L^2$ is implicit in the fluxes at the earth.

Using (\ref{fi}), we find the net flux of
electron neutrinos at the earth:
\barr
F_e & = & \sum_i |U_{ei}|^2  F_i \nonumber \\
 & = & F_e^0 \sum_i |U_{ei}|^2 a_i + 
                 F_x^0 ( 1 - \sum_i |U_{ei}|^2 a_i)~~,
\earr 
where we have taken into account the unitarity condition
$\sum |U_{ei}|^2 = 1$.
The final electron neutrino flux reaching the earth can thus
be written as
\beq
F_e = p F_e^0 + (1-p) F_x^0~~,
\label{flux-e}
\eeq
where
\barr
p & \equiv & \sum_i |U_{ei}|^2 a_i \nonumber \\
 & =  &
|U_{e1}|^2  P_{H} P_{L} + |U_{e2}|^2 (P_{H} - P_{H} P_{L})
+ |U_{e3}|^2 (1 - P_{H})~~.
\label{p-surv}
\earr
According to (\ref{flux-e}), $p$ may be interpreted as the 
total survival probability of electron neutrinos.

The original total flux of the neutrinos 
$\nu_e, \nu_\mu, \nu_\tau$
is $F_e^0 + 2 F_x^0$.
Using the conservation of flux, we find the combined flux of 
$\nu_{\mu}$ and $\nu_{\tau}$ at the earth 
($F_\mu + F_\tau$) as
\beq
F_\mu + F_\tau = (1-p) F_e^0 + (1 + p) F_x^0~~.
\label{fmutau}
\eeq

Note that the final fluxes of the flavor states at the earth
(\ref{flux-e},\ref{fmutau}) can be written in terms of only 
the survival probability $p$.
This is a consequence of two
facts: 1) at each transition, one of the neutrinos is 
decoupled so that the task reduces to $2\nu$ mixing, 
and 2) the original fluxes of $\nu_{\mu'}$ and
$\nu_{\tau'}$ are equal.

%%%%%%%%%%%%%%%%%%%%%%%%%%%%%%%%%%%%%%%%%%%%%%%%%%%%%%%%%%%%%%
\subsection{Conversion probabilities for antineutrinos}
%%%%%%%%%%%%%%%%%%%%%%%%%%%%%%%%%%%%%%%%%%%%%%%%%%%%%%%%%%%%%

Let us  consider the antineutrino transitions.
In the high matter density region 
($\rho \gg \rho_H, \rho_L$),
the antineutrino flavor eigenstates coincide with the
eigenstates in the medium as (see Figs.~\ref{generic}a,
\ref{generic}c):
\beq
\bar{\nu}_{1m} = \bar{\nu}_e ~,~
\bar{\nu}_{2m} = \bar{\nu}_{\mu'} ~,~
\bar{\nu}_{3m} = \bar{\nu}_{\tau'} ~~,
\eeq
so that the original fluxes of antineutrino
eigenstates in the medium equal
\beq
\bar{F}_{1m}^0 =   F_{\bar{e}}^0 ~~,~~
\bar{F}_{2m}^0  =   F_x^0 ~~,~~
\bar{F}_{3m}^0  =  F_x^0~~.
\eeq
The small mixing angle
$\theta_{e3}$ is further suppressed in the
medium, so the 
$\bar{\nu}_e \leftrightarrow \bar{\nu}_3$ transitions
are negligible.
The state $\bar{\nu}_{3m}$, being far from the
level crossings, propagates adiabatically:
$\bar{\nu}_{\tau'} \to \bar{\nu}_3$.
Depending on the parameters of the solution for the solar
neutrino problem, the propagation of the other two states
may be adiabatic or non-adiabatic \cite{spergel}. 
From considerations similar to those in the neutrino channel,
we get
\beq
F_{\bar{e}} =  \bar{p} F_{\bar{e}}^0 + (1-\bar{p}) F_x^0~~,
\label{flux-ebar}
\eeq
where $\bar{p}$, the survival probability of $\bar{\nu}_e$,
equals
\beq
\bar{p} = |U_{e1}|^2 (1 - \bar{P}_L) + |U_{e2}|^2 \bar{P}_L~~.
\label{pbar-expr}
\eeq
In the case of completely adiabatic propagation, $\bar{P}_L = 0$ and 
\beq
\bar{p} = |U_{e1}|^2.
\label{pbar-expr2}
\eeq

Let $F_{\bar{\mu}}+ F_{\bar{\tau}}$ be the
combined flux of $\bar{\nu}_\mu$ and $\bar{\nu}_\tau$.
From (\ref{flux-ebar}) and the conservation of flux,
we get the combined flux of non-electron antineutrinos:
\beq
F_{\bar{\mu}}+ F_{\bar{\tau}}  =  (1-\bar{p}) F_{\bar{e}}^0 +
        (1 + \bar{p}) F_x^0 ~~.
\label{fmutaubar}
\eeq

From (\ref{fmutau}) and (\ref{fmutaubar}),
the total flux of the non-electron neutrinos (including
antineutrinos) is
\barr
4 F_x & = & F_\mu + F_\tau + F_{\bar{\mu}}+ F_{\bar{\tau}} \nonumber \\
   & = & (1-p) F_e^0 + (2 + p + \bar{p}) F_x^0 +
         (1-\bar{p}) F_{\bar{e}}^0~~.
\label{flux-x}
\earr

Summarizing, the equations (\ref{flux-e},\ref{flux-ebar}, \ref{flux-x})
can be written in the compact notation
\beq
\left( \begin{array}{c}
F_e \\ F_{\bar{e}} \\ 4 F_x
\end{array} \right) =
\left( \begin{array}{ccc}
p & 0 & 1-p \\ 0 & \bar{p} & 1 - \bar{p} \\
1-p & 1 - \bar{p} & 2 + p + \bar{p}
\end{array} \right) 
\left( \begin{array}{c}
F_e^0 \\ F_{\bar{e}}^0 \\  F_x^0
\end{array} \right)~~.
\label{tr-matrix}
\eeq
The above general expression holds for both, the normal and 
the inverted mass hierarchy. The 
survival probabilities $p$ and $\bar{p}$ for the
normal hierarchy are given
in (\ref{p-surv}) and (\ref{pbar-expr}) respectively.
These probabilities depend
on the specific parameters of the masses and mixing scheme.

%%%%%%%%%%%%%%%%%%%%%%%%%%%%%%%%%%%%%%%%%%%%%%%%%%%%%%%%%%%%
\subsection{Earth matter effects on the $\nu_e$ spectrum}
\label{earth}
%%%%%%%%%%%%%%%%%%%%%%%%%%%%%%%%%%%%%%%%%%%%%%%%%%%%%%%%%%%%%

The neutrino trajectory  inside the earth before 
reaching the detector  depends on the direction of the 
supernova relative to the earth and the time of the day.
The comparison of signals from different detectors would
allows one to reveal the earth matter effects.
Also certain features of  the energy spectra can reveal 
the earth matter effect even from the  observations in  one detector.

The mass eigenstates arriving at the surface of the earth 
oscillate in the earth matter.
Let $P_{ie}$ be the probability
that a mass eigenstate $\nu_i$ entering the
earth reaches the detector as a $\nu_e$.
The flux of $\nu_e$ at the detector is
\beq
F_e^D  = \sum_i P_{ie} F_i ~~.
\eeq
Inserting $F_i$ from (\ref{fi}), we get 
\beq
F_e^D =   F_e^0 \sum a_i P_{ie} +
        F_x^0 (1 - \sum_i a_i P_{ie}) ~~,
\eeq
where $a_i$'s are as defined in (\ref{ai-def}), and we have
used the unitarity condition $\sum_i P_{ie} = 1$.
Thus, the $\nu_e$ flux at the detector can  be written as
\beq
F_e^D = p^D F_e^0 + (1 - p^D)  F_x^0 ~~,
\label{fed}
\eeq
with
\beq
p^D = \sum_i a_i P_{ie}~~.
\eeq
Comparing $F_e^D$ (\ref{fed}) with $F_e$ (\ref{flux-e}), we find 
the difference in the $\nu_e$ fluxes at the detector
due to the propagation in earth equals 
\beq
F_e^D - F_e =   (p^D - p) (F_e^0 - F_x^0)~~,  
\label{fed-fe}
\eeq
where $p$ is given in (\ref{p-surv}).
The earth matter effect can be quantified by
the difference of probabililty ($p^D - p$):
\beq
p^D - p = \sum_i a_i (P_{ie} - |U_{ei}|^2)~~.
\label{deltap}
\eeq
Using definitions in (\ref{ai-def}) and $\sum_i P_{ie} = 1$, we can write
explicitly
\beq
p^D - p =  P_H (P_{2e} - |U_{e2}|^2) ( 1 - 2 P_{L})
+ (P_{3e} - |U_{e3}|^2)(1 - P_H - P_H P_L)~~.
\label{deltap-exp}
\eeq
The second term in (\ref{deltap-exp}) can  be neglected.
Indeed, inside the earth, $\nu_3$ oscillates with a very small
depth:
\beq
P_{3e} - |U_{e3}|^2 \lsim 
\left(\frac{2 E V_{earth}}{\Delta m^2_{atm}}\right)
\sin^2 2 \theta_{e3}~~,
\label{p3e}
\eeq
where $V_{earth}$ is the effective potential of
$\nu_e$ in the earth.
For neutrino energies of 5 -- 50 MeV,
we have 
$2 E V_{earth}/ \Delta m^2_{atm} \lsim 10^{-2}$.
Moreover, 
$\sin^2 2\theta_{e3} \leq 0.1$,
so that 
\beq
P_{3e} - |U_{e3}|^2 \leq 10^{-3}~~.
\label{p3e-ue3}
\eeq
Finally, we can write
\beq
p^D - p \approx  P_H (P_{2e} - |U_{e2}|^2) ( 1 - 2 P_{L})~~.
\label{deltap-exp2}
\eeq
In general, when the signals from two detectors D1 and D2 are compared,
we get the difference of fluxes 
\beq
F_e^{D1} - F_e^{D2} \approx P_H \cdot (1 - 2 P_L) 
\cdot (P_{2e}^{(1)} - P_{2e}^{(2)}) 
\cdot (F_e^0 - F_x^0)~~.
\label{fdiff-nu}
\eeq
where $P_{2e}^{(1)}$ and  $P_{2e}^{(2)}$ are the 
$\nu_e \leftrightarrow \nu_2$ oscillation probabilities  for the 
detectors D1 and D2 correspondingly. 

According to  (\ref{fdiff-nu}) the earth matter effect is factorized:
it is proportional to the difference in the original 
$\nu_e$ and $\nu_x$ fluxes, 
the conversion factor
$P_H (1 - 2 P_L)$ inside the star, and
the difference of earth oscillation
probabilities $P_{2e}$ at the two detectors. 
Let us consider these factors separately.

1) $F_e^0 - F_x^0$: 
Since the $\nu_e$ spectrum is softer than the $\nu_x$
spectrum, and the luminosities of both the spectra are
similar in magnitude \cite{janka-lum},
the term $(F_e^0 - F_x^0)$ 
is positive at low energies and becomes negative
at higher energies where the $\nu_x$ flux overwhelms the
$\nu_e$ flux. Therefore, the earth effect has a different sign 
for low and high energies, and there exists a critical energy
$E_c$, such that $F_e^0(E_c) = F_x^0(E_c)$, where this change
of sign takes place. Since the cross section of the
neutrino interactions increases with energy, 
the  Earth effect is expected to be more significant 
at higher energies (if all the other factors are only weakly 
sensitive to the neutrino energy).

2) $P_H (1 - 2 P_L)$: 
This factor characterizes the neutrino conversions
inside the star. $P_H$ can be looked upon as a suppression factor
due to the conversions at the higher resonance.
Indeed,
in the limit of $U_{e3} \to 0$ and $P_H \to 1$, 
(\ref{deltap-exp2}) reduces to the expression
for the earth effects in the case of two neutrino
mixing :
\beq
[p^D - p]_{2 \nu}= (P_{2e} -  |U_{e2}|^2) ( 1 - 2 P_{L})
\label{d-n}
\eeq
which is equivalent to the one used in literature
\cite{daynight} in the context of day-night effect 
for solar neutrinos. Therefore  (\ref{deltap-exp2}) can be
looked upon as the earth matter effect due to the 
two neutrino mixing (\ref{d-n}) suppressed by a factor of $P_H$.
The mixing of the 
third neutrino thus plays a major role, making the 
expected earth effects in the case of supernova neutrinos
smaller than those expected in the case of solar 
neutrinos for the same mixing scheme and in the same energy range.
If the $H$-resonance is completely adiabatic, 
the earth effect vanishes:
all the $\nu_e$ produced
are converted to $\nu_3$ in the star, and the earth
matter effect on  $\nu_3$ is negligibly small (as we have
established through (\ref{p3e})).

3) $P_{2e}^{(1)} - P_{2e}^{(2)}$: 
If the neutrino trajectory  crosses only the mantle of the earth, 
one can use a constant density approximation which gives 
\beq
P_{2e}^{(1)} - P_{2e}^{(2)}  \approx \sin 2\theta_{e2}^m
~ \sin (2\theta_{e2}^m - 2\theta_{e2}) 
~\left[ \sin^2 \left( \frac{\pi d_1}{l_m} \right) -
\sin^2 \left( \frac{\pi d_2}{l_m} \right) \right]~~. 
\label{earth-nu}
\eeq
Here $\theta_{e2}^m$ and $l_m$ are  the mixing angle and 
oscillation length inside the earth respectively, and 
$d_i$ is the distance travelled by the
neutrinos inside the earth before reaching the detector $D_i$.
The first two terms on the right hand side of
(\ref{earth-nu}) are positive for the scenarios with the SMA and LMA
solutions, so that the 
sign of $(P_{2e}^{(1)} - P_{2e}^{(2)})$ is the same as
the sign of the term inside the square bracket in
(\ref{earth-nu}).
For the scenario with the VO solution, 
the earth matter effects are negligible: 
for $\Delta m^2 \sim 10^{-10}$ eV$^2$, the mixing
in the earth matter is highly suppressed, so that
$\sin 2\theta_{e2}^m$ in (\ref{earth-nu}) is very small.

If  neutrinos cross  both the mantle and the core, the 
parametric enhancement of oscillations may occur, which leads to
the appearance of parametric peaks apart from the 
peaks due to the MSW resonances in the core and the mantle
\cite{adls}. Correspondingly the factor $(P_{2e}^{(1)} - P_{2e}^{(2)})$ 
will be a more complicated function of the neutrino energy.

To summarize,
the earth matter effects on the $\nu_e$ spectrum can
be significant only for the scenarios with the SMA or LMA
or LOW solutions. Moreover, the $H$ resonance needs to be
non-adiabatic in the case of normal mass hierarchy. 

%%%%%%%%%%%%%%%%%%%%%%%%%%%%%%%%%%%%%%%%%%%%%%%%%%%%%%%%%%%%%%%%%%
\subsection{Earth matter effects on the $\bar{\nu}_e$ spectrum}
\label{earth-antin}
%%%%%%%%%%%%%%%%%%%%%%%%%%%%%%%%%%%%%%%%%%%%%%%%%%%%%%%%%%%%%%%%%%

The oscillation effects, are determined 
by  the survival probability of the
electron antineutrinos, $\bar{p}^D$, at the 
detector. Considerations similar to those for
(\ref{pbar-expr}) lead to
\beq
\bar{p}^D = \bar{P}_{1e} (1 - \bar{P}_L) +
\bar{P}_{2e} \bar{P}_L~~.
\label{pbar-d}
\eeq
Then  from Eqs. (\ref{flux-ebar}), (\ref{pbar-expr}) and (\ref{pbar-d}),
we obtain 
\beq
F^D_{\bar{e}} - F_{\bar{e}} =
(\bar{P}_{1e} - |U_{e1}|^2) \cdot (1 - 2 \bar{P}_L)
\cdot (F_{\bar{e}}^0 - F_x^0)~~,
\label{fd-f-ebar}
\eeq
where we have neglected the oscillations of $\bar{\nu}_3$ 
inside the earth. 
Generalizing the result in  (\ref{fd-f-ebar}), we find the difference in
the
fluxes at two detectors $D1$ and $D2$: 
\beq
F_{\bar{e}}^{D1} - F_{\bar{e}}^{D2} \approx
(\bar{P}_{1e}^{(1)} - \bar{P}_{1e}^{(2)}) \cdot (1 - 2 \bar{P}_L)
\cdot (F_{\bar{e}}^0 - F_x^0)~~.
\label{fdiff-nubar}
\eeq 
The earth matter effects for the antineutrinos are 
thus also factorized: they are proportional to the
difference in the original $\nu_e$ and $\nu_x$ fluxes,
the factor of $(1 - 2 \bar{P}_L)$ which depends on the
conversions inside the star, and the difference 
$(\bar{P}_{1e}^{(1)} - \bar{P}_{1e}^{(2)})$ between the
oscillation probabilities inside the Earth for the neutrinos reaching
the two detectors. Note that due to the absence of the 
$H$-resonance in the 
antineutrino channel, there is no suppression factor 
similar to  $P_H$ (\ref{fdiff-nu}) in the neutrino case. 

The factor $(F_{\bar{e}}^0 - F_x^0)$ is positive at low energies and
negative at high energies. The matter effects then change sign at
an energy $\bar{E}_c$ where $F_{\bar{e}}^0(\bar{E}_c) = 
F_x^0(\bar{E}_c)$. 
Since the neutrino cross section   increases with energy,  
the observed effect is expected to be larger 
at higher energies (if all the other factors are only weakly 
sensitive to the antineutrino energy). 

In the approximation of a constant density, the 
factor depending on the oscillation probabilities inside
the earth is
\beq
\bar{P}_{1e}^{(1)} - \bar{P}_{1e}^{(2)} \approx  
- \sin 2\bar{\theta}_{e2}^m
~\sin (2\bar{\theta}_{e2}^m - 2\theta_{e2}) 
~\left[ \sin^2 \left( \frac{\pi d_1}{l_m} \right) -
\sin^2 \left( \frac{\pi d_2}{l_m} \right) \right]~~,
\label{earth-nubar}
\eeq
where $\bar{\theta}_{e2}^m$ is the mixing angle inside the earth 
for the antineutrinos. For the antineutrino channel 
$\bar{\theta}_{e2}^m < \theta_{e2} \ll 1$ for SMA solution and 
$\bar{\theta}_{e2}^m$ is strongly suppressed by matter 
in the VO case.  
Therefore 
the earth matter effects on the
$\bar{\nu}_e$ spectrum can be significant only for the
scenario with the LMA (as well as  LOW) solution. In this scenario,
$\sin 2\bar{\theta}_{e2}^m >0$ and 
$\sin (2\bar{\theta}_{e2}^m - 2\theta_{e2}) <0$,
so that according to  (\ref{earth-nubar}), the sign of 
$(\bar{P}_{1e}^{(1)} - \bar{P}_{1e}^{(2)})$ is the
same as the sign of the oscillation term inside the square bracket
in (\ref{earth-nubar}).

To summarize, in the case of normal mass hierarchy,
the earth matter effects on the $\bar{\nu}_e$
spectrum are significant only for the scenario with the LMA
solution. The effects are practically unaffected by the mixing of
$\bar{\nu}_3$ and change sign at an energy $\bar{E}_c$ such that
$F_{\bar{e}}^0(\bar{E}_c) =
F_x^0(\bar{E}_c)$.

%%%%%%%%%%%%%%%%%%%%%%%%%%%%%%%%%%%%%%%%%%%%%%%%%%%%%%%%%%%%%%%%%%%%%%%
%%%%%%%%%%%%%%%%%%%%%%%%%%%%%%%%%%%%%%%%%%%%%%%%%%%%%%%%%%%%%%%%%%%%%%%
\section{Effects of neutrino conversion for  the mass spectra with normal
hierarchy}
\label{normal}
%%%%%%%%%%%%%%%%%%%%%%%%%%%%%%%%%%%%%%%%%%%%%%%%%%%%%%%%%%%%%%%%%%%%%%%%

In this section,
we shall consider the neutrino conversion effects 
for specific $3\nu$ schemes with the normal mass hierarchy.
The general expressions for the transition probabilities and 
the neutrino fluxes at the detectors are given in
Sec.~\ref{final-spec}, where they are expressed in terms of the 
total survival probabilities  $p$ and $\bar{p}$ of electron neutrinos
and electron antineutrinos respectively. The values of 
$p$ and $\bar{p}$ need to be calculated separately for
each specific scheme. 
Notice that, given a scheme, the least known parameter is
$|U_{e3}|^2$.  We shall consider the effects of neutrino
conversion in the three possible regions of $|U_{e3}|^2$:
these correspond to the regions I, II and  III 
as described in Sec.~\ref{conv} (see also 
Figs. \ref{pfvar} and \ref{e-dep}.) 

%%%%%%%%%%%%%%%%%%%%%%%%%%%%%%%%%%%%%%%%%%%%%%%%%%%%%%%%%%%%%%%%%%%
\subsection{The scheme with the SMA solution}
\label{sma}
%%%%%%%%%%%%%%%%%%%%%%%%%%%%%%%%%%%%%%%%%%%%%%%%%%%%%%%%%%%%%%%%%%%

\begin{figure}[htb]
\hbox to \hsize{\hfil\epsfxsize=10cm\epsfbox{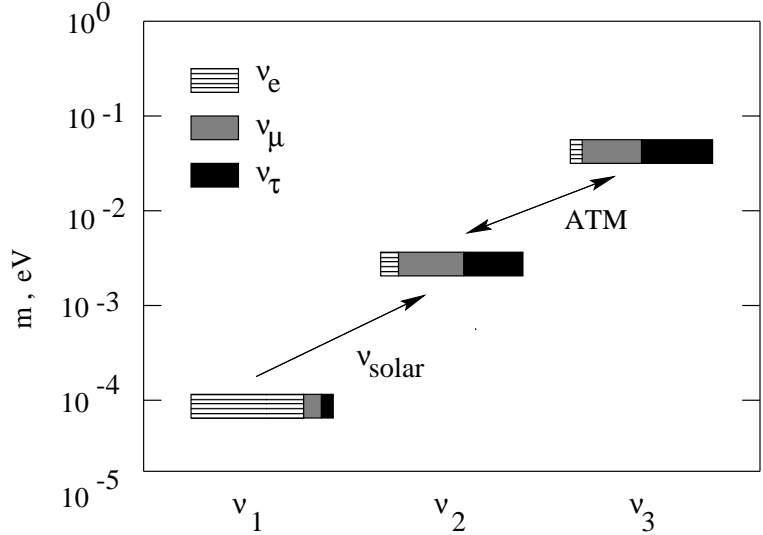}\hfil}
\caption{~~Neutrino mass and mixing
pattern for the scheme with the 
small mixing angle MSW solution of the solar neutrino problem.
The boxes correspond to the mass eigenstates. The sizes
of different regions in the boxes show admixtures
of different flavors.
Weakly hatched regions correspond to the electron
flavor, strongly hatched regions depict the muon flavor, black regions
represent the tau flavor.
\label{sma-box}}
\end{figure}

The mass and flavor spectrum of the scheme  is shown in 
Fig.~\ref{sma-box}.
The non-electron neutrinos $\nu_{\mu}$ and $\nu_{\tau}$ 
mix strongly
in the mass eigenstates $\nu_{2}$ and $\nu_{3}$.
The electron flavor is weakly mixed:
it  is mainly in $\nu_{1}$ with small admixtures
in the heavy states.
The solar neutrino data is explained via 
the small mixing angle MSW solution, with the parameters 
as given in (\ref{sma-param}). The level crossing scheme is
shown in Fig.~\ref{generic}a. Both the resonances
are in the neutrino channel.

In this scheme, 
\beq
|U_{e1}|^2 \approx 1 ~~,~~ |U_{e2}|^2
\approx \frac{1}{4} \sin^2 2\theta_\odot \sim 10^{-3}~~.
\label{sma-par}
\eeq

Let us first consider the antineutrino channels. 
There is no resonance here, and the mixing in matter is suppressed.  
As a consequence, the adiabaticity condition is fulfilled  
in the $L$ layer and  $\bar{P}_L \approx  0$. 
Then according to the level crossing scheme (Fig.~\ref{generic}a), the  
following transitions occur inside the star:
$$
\bar{\nu}_e \to \bar{\nu}_1 ~~,~~
\bar{\nu}_{\mu'} \to \bar{\nu}_2 ~~,~~
\bar{\nu}_{\tau'} \to \bar{\nu}_3~~. 
$$
Using $|U_{e2}|^2 \ll 1$ and $\bar{P}_L \approx  0$,  
we get from  (\ref{pbar-expr})
the survival probability for $\bar{\nu}_e$:  
\beq
\bar{p} \approx |U_{e1}|^2 ( 1 - \bar{P}_L) \approx 1~~.
\eeq

Let us consider now neutrino channels. 
The value of the $\nu_e$ survival probability 
$p$ depends on the region in which
the oscillation parameters of the  $H$ resonance lie.\\

{\bf Region I:} 
Since the $H$ resonance is adiabatic, 
the  level crossing scheme (Fig.~\ref{generic}a) leads to the
following transitions:
$$
\nu_e \to \nu_3~~,~~ \nu_{\mu'} \to \nu_1, \nu_2 ~~,~~
\nu_{\tau'} \to \nu_1, \nu_2~~.
$$
At $P_H \approx 0$, the Eq.  (\ref{p-surv})
gives 
\beq
p \approx |U_{e3}|^2 \leq 0.03~~.
\eeq
The flavor transitions are then complete, and
independent of the  adiabaticity at the $L$ resonance.

The final neutrino (and antineutrino) fluxes are
\beq
F_e \approx F_x^0~~,~~
F_{\bar{e}} \approx F_{\bar{e}}^0 ~~,
~~4 F_x \approx F_e^0  + 3 F_x^0~~.
\eeq
The neutrino spectra have the following features:
\begin{itemize}
\item
The neutronization $\nu_e$ peak disappears
(it is suppressed by a factor of $|U_{e3}|^2 \leq 0.03$).
Instead one would expect a $\nu_x$ neutronization peak
which may be detected by the neutral current interactions.

\item
The antineutrino signal is practically unchanged. 

\item
The $\nu_e$ has the hard spectrum of $F^0_x$,
and therefore
$$\langle E_{\nu_e} \rangle > 
\langle E_{\bar{\nu}_e} \rangle~~,$$ 
which is a 
clear signal of mixing.

\item
The $\nu_x$ spectrum is {\it composite, i.e.} it 
contains both the  soft (original $\nu_e$) and the hard 
(original $\nu_x$) components.

\end{itemize}
The earth matter effects on the $\nu_e$ spectrum (\ref{fdiff-nu})
are suppressed by the factor of $P_H \approx 0$.
Also, since the mixing in the antineutrino channel is
suppressed in matter, the earth matter effects on the
$\bar{\nu}_e$ spectrum are negligible.
As a result, one expects practically the same neutrino (and
antineutrino) signal
in all detectors.\\

{\bf Region II.} Neutrinos jump partially between  the third 
and the second levels in $H$ region, so that the following 
transitions occur:
$$
\nu_e \to \nu_1, \nu_2, \nu_3 ~~,~~
\nu_{\mu'} \to \nu_1, \nu_2 ~~,~~
\nu_{\tau'} \to \nu_1, \nu_2, \nu_3 ~~.
$$
The $\nu_e$ survival probability $p$ is given by
the general formula (\ref{p-surv}).
The value of $P_H$, and hence the survival
probability $p$, depends on the neutrino energy, although 
this   dependence is  relatively weak, 
as demonstrated in
Figs.~\ref{pfvar} and ~\ref{e-dep}.

The features of the spectra are: 
\begin{itemize}
\item the neutronization peak is distributed in 
both, $\nu_e$ and $\nu_x$,
\item 
the $\nu_e$ and $\nu_x$ spectra are composite,
\item the $\bar{\nu}_e$ spectrum is unchanged.
\end{itemize}
The earth matter effects can be observed in the $\nu_e$ spectrum.   
The magnitude of the effects is  proportional to $P_H$ (\ref{fdiff-nu}). 
For $\bar{\nu}_e$ spectrum the effect is  negligible. \\

{\bf Region III.} 
No conversion occurs at the $H$ resonance,  so that    
in the star
$$
\nu_e \to \nu_1, \nu_2 ~~,~~
\nu_{\mu'} \to \nu_1, \nu_2 ~~,~~
\nu_{\tau'} \to \nu_3~~.
$$
From (\ref{p-surv}) and (\ref{sma-par}) we get using $P_H \approx 1$:  
\beq
p \approx |U_{e1}|^2 P_L + |U_{e2}|^2 (1 - P_L) 
\approx P_L~~.
\eeq
The features of the spectra are the same as those described 
above in the case of region II.

\begin{figure}[htb]
\hbox to \hsize{\hfil\epsfxsize=8cm\epsfbox{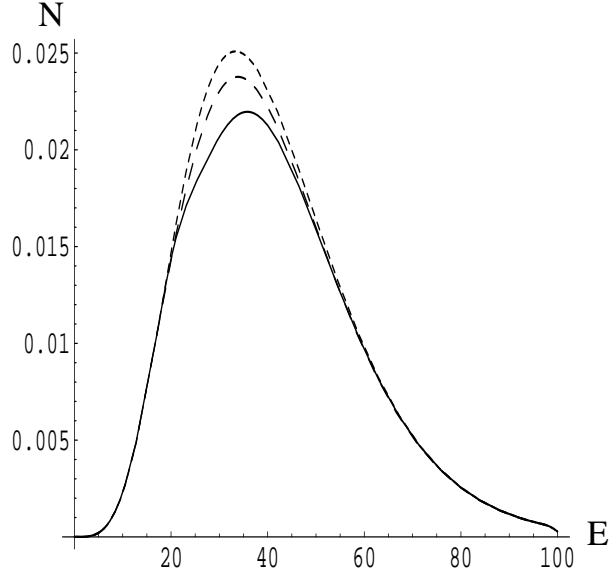}\hfil}
\caption{~~The earth matter effects on the $\nu_e$ spectrum
for $P_H=1$ and $P_L = 0$ in the scheme with the SMA solution
($\Delta m^2 = 10^{-5}$ eV$^2~,~ \sin^2 2 \theta_\odot = 10^{-2}$).
The dotted, dashed and solid lines 
show the number of $\nu_e-N$ charged current events
when the distance travelled by the neutrinos
through the mantle of the earth is
$d=0$ km, $d=6000$ km, and $d=10000$ km respectively.
\label{earth-sma}}
\end{figure}

The earth matter effects on the $\nu_e$ spectrum
in the SMA scheme can be significant
in regions II and III, where the factor of $P_H$
is of the order of unity.
Indeed, in (\ref{earth-nu}),
the value of $(P_{2e} - |U_{e2}|^2)$ can be as large
as 0.25 in the energy range of 20-40 MeV, where
the term $(F_e^0 - F_x^0)$ is also significant.

In Fig.~\ref{earth-sma}, we show the $\nu_e$ 
spectrum for different distances travelled by the neutrinos
through the earth. 
As follows from Fig.~\ref{earth-sma}, the net effect 
is $\lsim 10$\% even with optimistic values of
$P_H, P_L$ and $\theta_{e2}$.

%%%%%%%%%%%%%%%%%%%%%%%%%%%%%%%%%%%%%%%%%%%%%%%%%%%%%%%%%%%%%%%%%%%%%%
\subsection{The scheme with the LMA solution}
\label{lma}
%%%%%%%%%%%%%%%%%%%%%%%%%%%%%%%%%%%%%%%%%%%%%%%%%%%%%%%%%%%%%%%%%%%%%%%

\begin{figure}[htb]
\hbox to \hsize{\hfil\epsfxsize=10cm\epsfbox{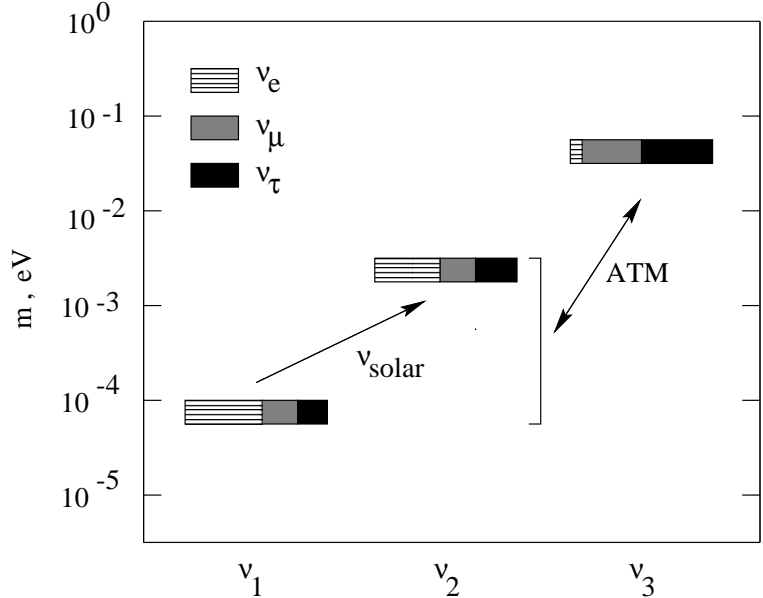}\hfil}
\caption{~~Neutrino mass and mixing
pattern  of  the scheme with the large mixing angle 
MSW solution for the solar
neutrino problem. 
\label{lma-box}}
\end{figure}

The mass and flavor spectrum of the scheme
is shown in
Fig.~\ref{lma-box}.
The solar neutrino data is explained via 
the $\nu_e \rightarrow \nu_{2}$
resonant conversion inside the sun with a large
vacuum mixing angle (\ref{lma-param}).
The level crossing scheme is shown in
Fig.~\ref{generic}c.
It differs from the previous one by the mixing at the
$L$-resonance.
Since the vacuum mixing angle $\theta_\odot$ is
large, there is a significant 
$\bar{\nu}_e \leftrightarrow \bar{\nu}_2$ conversion
even though the antineutrinos do not encounter any
resonances.
The evolution is adiabatic in the antineutrino channel
in both the $H$ and $L$ layers.
The following transitions occur:
$$
\bar{\nu}_e \to \bar{\nu}_1~~,~~
\bar{\nu}_{\mu'} \to \bar{\nu}_2~~,~~
\bar{\nu}_{\tau'} \to \bar{\nu}_3~~.
$$
Using $\bar{P}_L \approx 0$, the $\bar{\nu}_e$
survival probability (\ref{pbar-expr}) equals
\beq
\bar{p} \approx |U_{e1}|^2 \approx \cos^2  \theta_\odot~~.
\label{lma-ebar}
\eeq

Let us consider transitions in the neutrino channels. 
The $L$-resonance is adiabatic
(Fig.~\ref{adia-box}),
$P_L \approx 0$, and from (\ref{p-surv}),
we get the $\nu_e$- survival probability
\barr
p & \approx & |U_{e2}|^2 P_H + |U_{e3}|^2 (1 - P_H) \nonumber \\
 & \approx & \sin^2 \theta_\odot P_H + |U_{e3}|^2 
(1 - P_H)~~.
\label{lma-p}
\earr
Notice that, depending on the value of $P_H$,   
the $\nu_e$- survival probability takes the values between
\beq
|U_{e3}|^2 \leq p \leq |U_{e2}|^2~~.
\eeq

The  fluxes at the earth are determined by
(\ref{tr-matrix}), with $p$ and $\bar{p}$ given
in (\ref{lma-p}) and (\ref{lma-ebar}).
Their features  depend on
the adiabaticity at the $H$ resonance, which is
decided by the region in which $|U_{e3}|^2$ lies. \\

{\bf Region I.} The level crossing scheme (Fig.~\ref{generic}c) 
with adiabatic transitions in 
the $H$ resonance layer leads to the following 
transitions inside the star: 
$$
\nu_e \to \nu_3 ~~,~~
\nu_{\mu'} \to \nu_1~~,~~
\nu_{\tau'} \to \nu_2 ~~.
$$
Since $P_H \approx  0$ we get  $p \approx |U_{e3}|^2$, 
and 
the final spectra at the earth: 
\barr
F_e & \approx & |U_{e3}|^2 F_e^0 + ( 1 - |U_{e3}|^2) F_x^0 
\approx F_x^0 ~~,\nonumber \\
F_{\bar{e}} & \approx & \cos^2 \theta_\odot F_{\bar{e}}^0 +
\sin^2 \theta_\odot F_x^0 ~~,\nonumber \\
4 F_x & \approx & F_e^0 + \sin^2 \theta_\odot F_{\bar{e}}^0 +
(2 + \cos^2 \theta_\odot) F_x^0~~.
\label{lma-I}
\earr
The following features can be observed:

\begin{itemize}
\item
The neutronization peak almost disappears from
the $\nu_e$ channel 
(suppressed by a factor of $|U_{e3}|^2 \lsim 0.03 $) and  
appears in the $\nu_x$ channel.

\item
The $\bar{\nu}_e$ spectrum is composite: 
A mixture of the original $\bar{\nu}_e$
spectrum and the original harder $\nu_x$ spectrum,
is  determined by the solar neutrino
mixing angle. This feature distinguishes the LMA scheme  from the
one with the SMA solution.

\item
The $\nu_e$ spectrum is hard, practically coinciding with 
the original $\nu_x$ spectrum. This leads to
$$
\langle E_{\nu_e} \rangle > \langle E_{\bar{\nu}_e} \rangle~~.
$$

\item
The $\nu_x$ spectrum contains 
components of all the three original spectra
($\nu_e, \bar{\nu}_e, \nu_x$).

\end{itemize}
No earth matter effects are  expected in the $\nu_e$
spectrum, since $P_H \approx 0$. At the same time,
 the $\bar{\nu}_e$
spectrum can show significant earth matter effects.
This is an important signature of the scenario.\\

{\bf Region II.} The jump probability in H resonance is substantial 
%$P_H \sim 0.1 - 0.9$. 
and the following transitions occur:
$$
\nu_e \to \nu_2, \nu_3 ~~,~~
\nu_{\mu'} \to \nu_1 ~~,~~
\nu_{\tau'} \to \nu_2, \nu_3~~.
$$
The final spectra are determined by (\ref{tr-matrix})
with $p$ given by (\ref{lma-p}):  
$$
p \approx \sin^2 \theta_\odot~ P_H~~.
$$
Depending on  $P_H$ 
and $\sin^2 \theta_\odot \approx |U_{e2}|^2 \approx 0.2-0.4$,  
the probability $p$  can take values between 
0.02 and 0.4. 

The final spectra can be written as
\barr
F_e & \approx & \sin^2 \theta_\odot P_H F_e^0 +
(1 - \sin^2 \theta_\odot P_H) F_x^0 \nonumber \\
F_{\bar{e}} & \approx & \cos^2 \theta_\odot F_{\bar{e}}^0 +
\sin^2 \theta_\odot F_x^0 \nonumber \\
4 F_x & \approx &
(1-\sin^2 \theta_\odot P_H) F_e^0 + \sin^2 \theta_\odot F_{\bar{e}}^0 +
(3-\sin^2 \theta_\odot+\sin^2 \theta_\odot P_H) F_x^0~~.
\label{lma-III}
\earr
The features of the spectra are:
\begin{itemize}

\item The neutronization peak consists of both, $\nu_e$
and $\nu_x$ fluxes, with the proportions
determined by $\theta_\odot$ and $P_H$ . 

\item
The $\nu_e$ and $\bar{\nu}_e$ spectra are composite,
with admixture of the hard component  determined
by the solar mixing angle. 

\item 
The $\nu_x$ spectrum contains 
components of all the three original spectra
($\nu_e, \bar{\nu}_e, \nu_x$).

\end{itemize}
The earth matter effects can be significant for 
both the $\nu_e$ and $\bar{\nu}_e$ spectra.
The effects on the $\nu_e$ spectrum are suppressed 
moderately by a factor of $P_H$.\\

{\bf Region III:}
Here $P_H \approx 1$, so that according to (\ref{lma-p}),
$p \approx |U_{e2}|^2 \approx 
0.2 - 0.4$.
The $H$-resonance is 
inoperative, and the following transitions occur:
$$
\nu_e \to \nu_2 ~~,~~ \nu_{\mu'} \to \nu_1 ~~,~~
\nu_{\tau'} \to \nu_3~~. 
$$
The features of the spectra are
the same as those described in the case of the region II.
Since $P_H \approx 1$, the earth matter effects are 
expected to be significant for both $\nu_e$ and 
$\bar{\nu}_e$ spectra.\\

Let us consider in some details the 
the earth matter effects which can give important signature of the 
scheme under consideration. For  the $\nu_e$ spectrum  
the effect  can be significant in regions II and III.
Inserting $P_L = 0$ in (\ref{earth-nu}), 
we get  
\beq
F_e^D - F_e \approx P_H \cdot (P_{2e}-|U_{e2}|^2)
\cdot  (F_e^0 - F_x^0)~~
\label{fd-f-lma-nu}
\eeq
Using the constant density earth approximation (\ref{earth-nu})
we estimate that the factor
$(P_{2e} - |U_{e2}|^2)$ can be as large as 0.3
in the energy range of 20 -- 50 MeV. 

Similarly, for the antineutrinos, we get
\beq
F_{\bar{e}}^D - F_{\bar{e}} \approx  
(\bar{P}_{1e}-|U_{e1}|^2)
\cdot  (F_{\bar{e}}^0 - F_x^0)~~.
\label{fd-f-lma-nubar}
\eeq
In the antineutrino channel, 
the effect can be significant in all three
regions of $U_{e3}$, since there is no $P_H$ suppression,  and
again the factor $(\bar{P}_{1e} - |U_{e1}|^2)$
can be as large as 0.3  for  energies
around 20--50 MeV.  

\begin{figure}[htb]
\hbox to \hsize{\hfil\epsfxsize=14cm\epsfbox{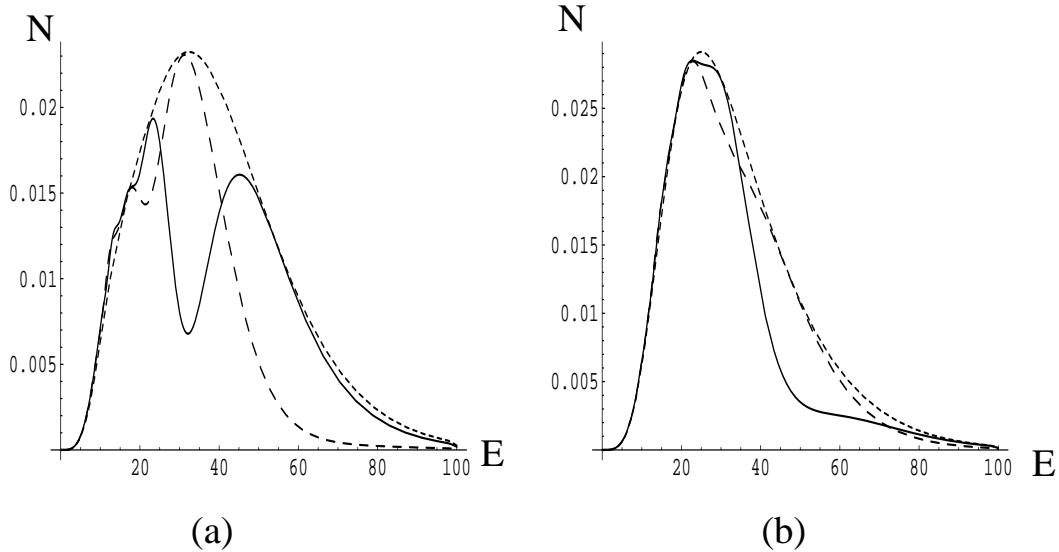}\hfil}
\caption{~~The earth matter effects on (a) the $\nu_e$ spectrum
and (b) the $\bar{\nu}_e$ spectrum
for $P_H=1$ in the scheme with the LMA solution
($\Delta m^2 =2 \cdot 10^{-5}$ eV$^2,~\sin^2 2\theta_\odot= 0.9$).
The dotted, dashed and solid lines 
show the spectra of the number of $\nu-N$ charged current events 
when the distance travelled by the neutrinos
through the earth is
$d=0$ km, $d=4000$ km, and $d=6000$ km respectively. 
\label{earth-lma}}
\end{figure}

In Fig.~\ref{earth-lma}, we show the $\nu_e$ and $\bar{\nu}_e$
spectra for different distances travelled by the neutrinos
through the earth. 
The sign of the earth effect (\ref{fd-f-lma-nu}, 
\ref{fd-f-lma-nubar}) is determined by the sign of
$(F_e^0 - F_x^0)$ in the neutrino channel and 
$(F_{\bar{e}}^0 - F_x^0)$ in the antineutrino channel.
The effect changes sign at $E = E_c$ ($E = \bar{E}_c$)
for the neutrinos (antineutrinos),
as described in Sec.~\ref{earth} (Sec.~\ref{earth-antin}).
Since $P_{2e} \geq |U_{e2}|^2$ and $\bar{P}_{1e} \geq |U_{e1}|^2$,
(\ref{fd-f-lma-nu}) and (\ref{fd-f-lma-nubar}) imply that
the earth effect results in the enhancement of
neutrino (antineutrino) signal for $E < E_c$ ($E<\bar{E}_c$)
and the depletion of the signal for 
$E > E_c$ ($E>\bar{E}_c$).
Since the cross section of the neutrino detection interactions is
larger at higher energies, a significant reduction
in the $\nu_e$ ($\bar{\nu}_e$) flux is expected at
energies greater than $E_c$ ($\bar{E}_c$).
It may manifest itself as a dip in the final spectrum
(see Fig.~\ref{earth-lma}),
and this spectral distortion can indicate 
the presence of significant earth matter effects,
even with the spectrum observed in only one detector.

%%%%%%% 4.3..%%%%%%%%%%%%%%%%%%%%%%%%%%%%%%%%%%%%%%%%%%%%%%%%%%%%%%%%%
\subsection{The scheme with the VO solution}
\label{vo}
%%%%%%%%%%%%%%%%%%%%%%%%%%%%%%%%%%%%%%%%%%%%%%%%%%%%%%%%%%%%%%%%

\begin{figure}[htb]
\hbox to \hsize{\hfil\epsfxsize=10cm\epsfbox{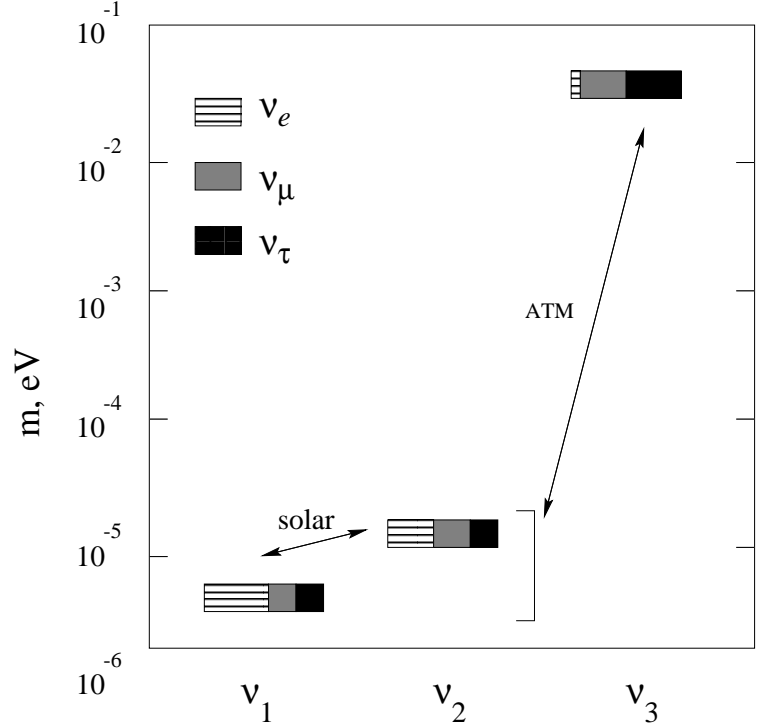}\hfil}
\caption{~~Neutrino mass and mixing
pattern  of  the scheme with the vacuum osacillation
 solution for the solar neutrino data.
\label{vo-box}}
\end{figure}

The mass and flavor spectrum for the scheme
is shown in Fig.~\ref{vo-box}.
The solar neutrino data is explained via
the vacuum oscillations with 
$\Delta m^2  \sim 10^{-10}$ eV$^2$ and
a large mixing between
$\nu_e$ and a combination of $\nu_\mu$ and 
$\nu_\tau$ (\ref{vo-param}).

The level crossing scheme is shown 
in Fig.~\ref{generic}c,
where the resonance $L$ is close to $\rho = 0$.
It may occur in the neutrino or antineutrino channel.
Depending on the details of the density profile
in the star at low densities, 
it lies either in the non-adiabatic region
(so that $P_L, \bar{P}_L \approx 1$) or in the transition region.

Let us first consider the antineutrino channel.
According to the level crossing scheme, we get the transitions  
$$
\bar{\nu}_e \to \bar{\nu}_1, \bar{\nu}_2~~,~~
\bar{\nu}_{\mu'} \to \bar{\nu}_1, \bar{\nu}_2~~,~~
\bar{\nu}_{\tau'} \to \bar{\nu}_3~~.
$$
The final fluxes are given by (\ref{p-surv}) and for  
$\bar{p}$ we get from (\ref{pbar-expr}) 
\beq
\bar{p} = \cos^2 \theta_\odot (1 - \bar{P}_L) +
\sin^2 \theta_\odot \bar{P}_L~~.
\eeq
This implies that the value of $\bar{p}$ lies
between   $\cos^2 \theta_\odot$ and
$\sin^2 \theta_\odot$, and  the $\bar{\nu}_e$
spectrum is composite.

The fluxes in the neutrino channel depend on the
region in which the oscillation parameters of the
$H$ resonance lie. \\

{\bf Region I.} The  $H$ resonance is adiabatic, and   
the following transitions occur in the neutrino
channel:
$$
\nu_e \to \nu_3~~,~~
\nu_{\mu'} \to \nu_1, \nu_2 ~~,~~
\nu_{\tau'} \to \nu_1, \nu_2~~.
$$
Since $P_H \approx 0$, 
from (\ref{p-surv}) we get 
$$p \approx |U_{e3}|^2 \leq 0.03~~.$$
The $\nu_e$ then has the hard spectrum of the original $\nu_x$.
The observed features of the final fluxes are the same as those
in the scheme with the LMA solution and 
an adiabatic $H$ resonance (\ref{lma-I}).
The earth matter effects are small 
for both the
$\nu_e$ and $\bar{\nu}_e$ spectra, since the mixing
angle $\bar{\theta}_{e2}$ is suppressed in the
earth matter.\\

{\bf Region II.} Jump probability 
in the $H$ resonance is substantial, so
that the transitions inside the star are:
$$
\nu_e \to \nu_1,\nu_2,\nu_3~~,~~
\nu_{\mu'} \to \nu_1, \nu_2 ~~,~~
\nu_{\tau'} \to \nu_1, \nu_2,\nu_3~~.
$$
From (\ref{p-surv}), we can infer
\beq
|U_{e3}|^2 \leq p
\leq \cos^2 \theta_\odot~~.
\eeq
where the lower bound corresponds to $P_H = 0$, whereas 
the upper bound is for $P_H = P_L = 1$. 
The $\nu_e$ spectrum is then composite.

The features of the spectra are:
\begin{itemize}

\item The neutronization peak should be  observed both in  $\nu_e$
and in $\nu_x$ channels.
\item
The $\bar{\nu}_e$ spectrum is composite,
the admixture of the original hard component is determined
by the solar mixing angle as well as by the degree of  adiabaticity
in the $H$ and $L$ conversion regions.
\item
The $\nu_e$ spectrum is composite,
the admixture of the original hard component is determined
by the solar mixing angle as well as the adiabaticity
in the $L$ resonance region.
\item
The $\nu_x$ spectrum contains
components of all the three original spectra
($\nu_e, \bar{\nu}_e, \nu_x$).
\end{itemize}
The earth matter effects  are small for both
the $\nu_e$ and the $\bar{\nu}_e$ spectra.\\

{\bf Region III.}
Since the $H$ resonance is completely non-adiabatic, 
the neutrino transitions take place in $L$ resonance region only:  
$$
\nu_e \to \nu_1,\nu_2~~,~~
\nu_{\mu'} \to \nu_1, \nu_2 ~~,~~
\nu_{\tau'} \to \nu_3~~.
$$
From (\ref{p-surv}) we get for  $P_H \approx 1$:  
\beq
p \approx |U_{e1}|^2 P_L + |U_{e2}|^2 (1- P_L)
\approx \cos^2 \theta_\odot P_L + \sin^2 \theta_\odot (1-P_L)~~.
\label{p-vo}
\eeq
Thus,
\beq
\sin^2 \theta_\odot \leq p \leq \cos^2 \theta_\odot~~.
\label{p-vo2}
\eeq
The $\nu_e$ - spectrum is composite. The features of the spectra
are the same as those observed above in region II.
The earth matter effects  are small for both
the $\nu_e$ and the $\bar{\nu}_e$ spectra.\\

\subsubsection*{The bi-maximal mixing scheme}

\begin{figure}[htb]
\hbox to \hsize{\hfil\epsfxsize=10cm\epsfbox{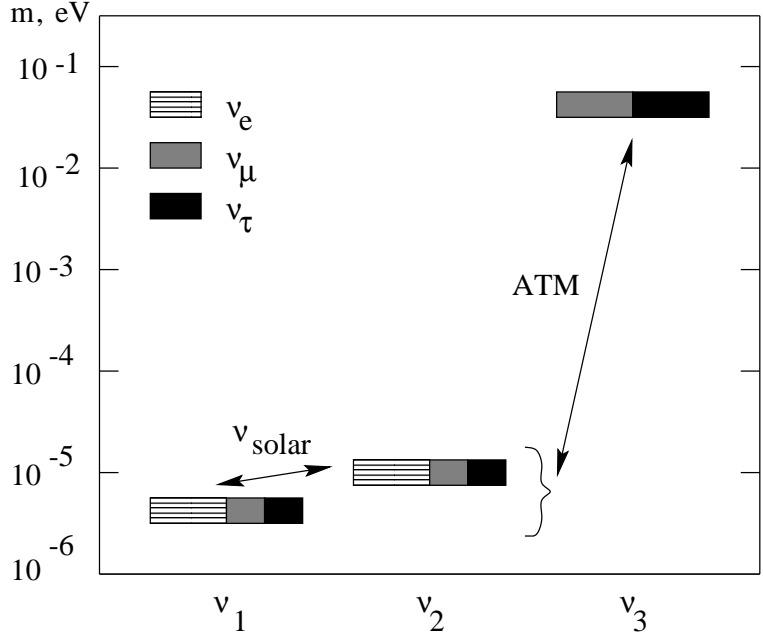}\hfil}
\caption{~~The neutrino mass and mixing pattern  of the  
``strict'' bimaximal
mixing  scheme.
\label{bimax-box}}
\end{figure}

An extreme case of the vacuum oscillations is the 
bimaximal mixing scenario \cite{bimaxim}. 
The mass and flavor spectrum is as shown
in Fig.~\ref{bimax-box}. 
$\nu_{\mu}$ and $\nu_{\tau}$ mix maximally in
$\nu_{\tau'} = (\nu_{\mu} + \nu_{\tau})/\sqrt{2}$; in turn,  the
orthogonal combination $\nu_{\mu'} \equiv (\nu_{\mu} -
\nu_{\tau})/\sqrt{2}$
maximally mixes maximally with
$\nu_e$.  In this scenario, $U_{e3} = 0$. The mass eigenstates are
then
$$
\nu_1 = (\nu_e -\nu_{\mu'})/\sqrt{2} ~~,~~
\nu_2 = (\nu_e +\nu_{\mu'})/\sqrt{2} ~~,~~
\nu_3 = \nu_{\tau'}~~.
$$
The following  transitions take place:
$$
\nu_e \to \nu_1, \nu_2 ~~,~~
\nu_{\mu'} \to \nu_1, \nu_2~~,~~
\nu_{\tau'} \to \nu_3~~,
$$
$$
\bar{\nu}_e \to \bar{\nu}_1, \bar{\nu}_2 ~~,~~
\bar{\nu}_{\mu'} \to \bar{\nu}_1, \bar{\nu}_2~~,~~
\bar{\nu}_{\tau'} \to \bar{\nu}_3~~.
$$
The mass eigenstates
$\nu_1$ and $\nu_2$ ($\bar{\nu}_1$  and $\bar{\nu}_2$) 
arriving at the earth are observed with
equal probability as electron or non-electron neutrinos.
The expected fluxes are then
\barr
F_e & = & (1/2) F_e^0 + (1/2) F_x^0~~, \nonumber \\
F_{\bar{e}} & = & (1/2) F_{\bar{e}}^0  + (1/2) F_x^0~~\nonumber \\
4 F_x & = & (1/2) F_e^0 + 3 F_x^0 + (1/2) F_{\bar{e}}^0 ~~.
\earr
The neutrino spectra are affected in the following way:
\begin{itemize}

\item
The neutronization peak consists of
equal fractions of electron and non-electron
neutrinos.

\item
The $\bar{\nu}_e$-spectrum  is ``composite'',
it has both
the soft and the hard components
in equal proportions.
\item
The $\nu_e$-spectrum is also composite with both
the soft
and  the hard components
in equal portions. This spectrum has
a lower average energy than the $\bar{\nu}_e$ spectrum,
but the high energy tails of both the spectra are identical.

\item
The $\nu_x$ spectrum has a mixture of all the
three initial spectra.

\end{itemize}
The earth matter effects
are small for both $\nu_e$ and $\bar{\nu}_e$ spectra
due to the small value of $\Delta m^2_\odot \sim 10^{-10}$
eV$^2$. 

%%%%%%%%%%%%%%%%%%%%%%%%%%%%%%%%%%%%%%%%%%%%%%%%%%%%%%%%%%%%%%%%%%%%%%%%
\section{Conversion probabilities and neutrino fluxes with
the inverted mass hierarchy}
\label{inverted}
%%%%%%%%%%%%%%%%%%%%%%%%%%%%%%%%%%%%%%%%%%%%%%%%%%%%%%%%%%%%%%%%%%%%%%%%

\begin{figure}[htb]
\hbox to \hsize{\hfil\epsfxsize=10cm\epsfbox{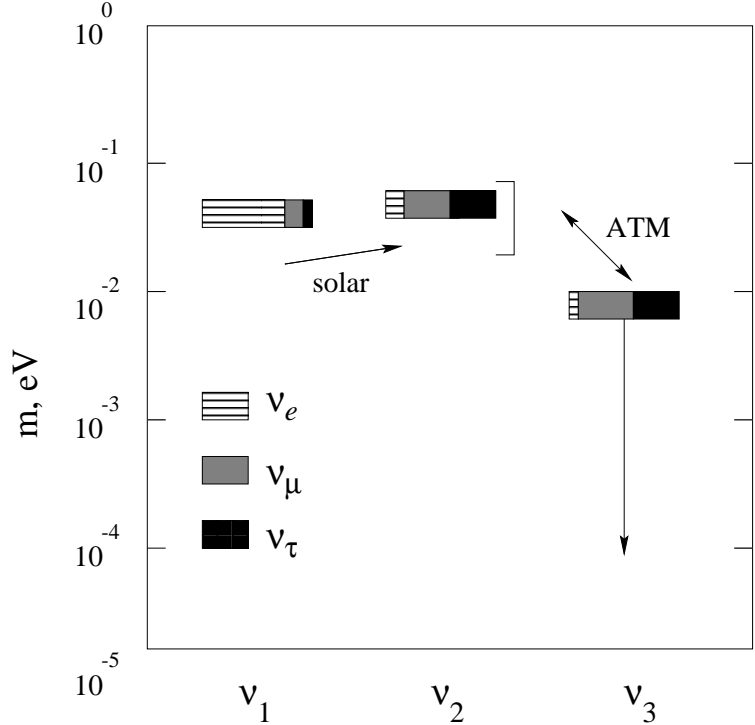}\hfil}
\caption{~~The neutrino mass and mixing pattern for the
scenario with inverted mass hierarchy and the SMA solution 
for the solar neutrino problem.
\label{inv-box}}
\end{figure}

In the case of the inverted mass hierarchy (\ref{inv-h}),
the mass eigenstates
$\nu_1$ and $\nu_2$ are heavy and degenerate, whereas the
third state $\nu_3$ is much lighter: $m_1 \approx m_2 \gg m_3$. 
(see e.g. Fig.~\ref{inv-box}. for the scenario with the SMA solution).

As shown in Figs.~\ref{generic}b and \ref{generic}d, 
the $H$-resonance is  in the antineutrino
channel, whereas
the $L$-resonance lies in the neutrino channel
for the SMA and LMA solutions.
In the case of the VO solution, the $L$ resonance can be in
either the neutrino or the antineutrino channel.

Like in the case of the normal mass hierarchy, 
the fluxes at the earth can
be written in terms of the 
survival probabilities $p$ and $\bar{p}$ of 
$\nu_e$ and $\bar{\nu}_e$ respectively, as given in
(\ref{tr-matrix}).
Let us find the  expressions for the
survival probabilities $p$ and $\bar{p}$.

As follows from the level crossing scheme
(Figs.~\ref{generic}a and \ref{generic}d), 
the matter eigenstates in the high density region
($\rho \gg \rho_H, \rho_L$) are 
\beq
\nu_{1m} = \nu_{\mu'} ~,~
\nu_{2m} = \nu_{e} ~,~
\nu_{3m} = \nu_{\tau'}~,
\eeq
\beq
\bar{\nu}_{1m} = \bar{\nu}_{\tau'} ~,~
\bar{\nu}_{2m} = \bar{\nu}_{\mu'} ~,~
\bar{\nu}_{3m} = \bar{\nu}_{e} ~~.
\eeq
The state $\nu_e$, 
which coincides with  $\nu_{2m}$ in
the production region,
crosses the $H$-resonance layer adiabatically
(since the resonance is in the antineutrino channel and the mixing 
in the neutrino channel is small),
and reaches the $L$-layer as $\nu_{2m}$.
It reaches the surface of the earth as $\nu_1$ with the
probability $P_L$ and as $\nu_2$ with the
probability $( 1- P_L)$. Then
\beq
p =  |U_{e1}|^2 P_L +  |U_{e2}|^2 (1 - P_L)~~.
\label{p-inv}
\eeq

The state $\bar{\nu}_e$, 
which coincides with $\bar{\nu}_{3m}$
in the production region, flips to $\bar{\nu}_{1m}$
in the resonance region $H$  with the probability $\bar{P}_L$.
Depending on the parameters of the solar neutrino solution, 
the  propagation of the antineutrinos in the $L$-resonance region 
({\it i. e.} near $\rho =0$) may or may not be adiabatic which is 
described by $\bar{P}_L$. Computing the probabilities of transitions  
of $\bar{\nu}_e$ to $\bar{\nu}_1$,  $\bar{\nu}_2$,   $\bar{\nu}_3$ and
performing projection of the result back onto $\bar{\nu}_e$ back we get 
(similarly to (\ref{p-surv}) 
the survival probability $\bar{p}$ for the $\bar{\nu}_e$:  
\beq
\bar{p} =  |U_{e1}|^2 \bar{P}_H (1 - \bar{P}_L) + 
|U_{e2}|^2 \bar{P}_H \bar{P}_L + |U_{e3}|^2 (1 - \bar{P}_H)~~.
\label{pbar-inv}
\eeq
(Here we have also averaged out the interference effects 
between the mass eigenstates, so that the three terms in 
(\ref{pbar-inv}) correspond to the three mass eigenstates.)\\

Let us consider the earth matter effects on the
$\nu_e$ and $\bar{\nu}_e$ spectra for the 
inverted mass hierarchy.
Since there is no $H$-resonance in the neutrino channel, 
the earth effects for the inverted hierarchy 
are the same as those described for the normal mass hierarchy
(see Sec.~\ref{earth}) with  $P_H = 1$:  
\beq
F_e^{D1} - F_e^{D2} \approx (P_{2e}^{(1)} - P_{2e}^{(2)}) \cdot
(1 - 2 P_L) \cdot (F_e^0 - F_x^0)~~.
\label{fdiff-nu-inv}
\eeq
The matter effects are
significant for the scheme with the LMA solution. The
suppression factor of $P_H$ [that was present for the
normal mass hierarchy - see eq.~(\ref{fdiff-nu})] is now absent, so the
earth matter effects can be larger than in the case of the normal 
mass hierarchy
(see Fig.~\ref{earth-lma}a).

The earth matter effects on the antineutrino spectra can be 
calculated as follows.
Using the same arguments as for (\ref{pbar-inv}),
we find the survival probability of $\bar{\nu}_e$ at 
the detector:
\beq
\bar{p}^D =  \bar{P}_{1e} \bar{P}_H (1 - \bar{P}_L) + 
\bar{P}_{2e} \bar{P}_H \bar{P}_L + \bar{P}_{3e} (1 - \bar{P}_H)~~.
\label{qd-inv}
\eeq 
Then (\ref{pbar-inv}) and (\ref{qd-inv}) give
\beq
\bar{p}^D - \bar{p} =
(\bar{P}_{1e}- |U_{e1}|^2) (1 - 2 \bar{P}_L) \bar{P}_H +
(\bar{P}_{3e} -|U_{e3}|^2) (1 - \bar{P}_H - \bar{P}_H \bar{P}_L)~~,
\label{del-pbar}
\eeq
where we have used $\sum \bar{P}_{ie} = 1$.
Since $\bar{\nu}_3$ oscillates inside the earth with a very
small depth [inequality (\ref{p3e-ue3}) is valid with $P_{3e}$
replaced by $\bar{P}_{3e}$],
the second term in (\ref{del-pbar}) can be neglected.
Therefore, finally we get
\beq
F_{\bar{e}}^{D1} - F_{\bar{e}}^{D2} \approx \bar{P}_H
(\bar{P}_{1e}^{(1)} - \bar{P}_{1e}^{(2)}) \cdot (1 - 2 \bar{P}_L) 
\cdot (F_{\bar{e}}^0 - F_x^0)~~.
\label{fdiff-nubar-inv}
\eeq

The earth matter effects on the $\bar{\nu}_e$ spectrum
for the inverted hierarchy 
are then the same as those described for the normal hierarchy 
in Sec.~\ref{earth},
but further suppressed by a factor of $\bar{P}_H$. 
Significant effects can be observed only with the LMA
scenario and when the $H$ resonance is non-adiabatic.

The earth matter effects on both $\nu_e$ and $\bar{\nu}_e$
spectra are factorized, as in the case of the normal mass hierarchy.
They are proportional to (i) the difference in original
$\nu_e$ ($\bar{\nu}_e$) and $\nu_x$ fluxes, (ii) a
factor which determines  the conversion inside the star, and
(iii) the difference of  the oscillation probabilities
inside the earth. 

Notice that the  expressions
for the earth matter effects in inverted mass hierarchy case 
(\ref{fdiff-nu-inv}, \ref{fdiff-nubar-inv}) have the same form as the 
expressions in the normal 
hierarchy case (\ref{fdiff-nu}, \ref{fdiff-nubar})
with substitution: $\nu \leftrightarrow \bar{\nu}$, 
$P_H \leftrightarrow \bar{P}_H$.

Let us now calculate the values of the survival probabilities
$p$ and $\bar{p}$ for the three scenarios with the
SMA, LMA and VO solutions for the solar neutrino problem
respectively. Within each scenario, the final spectra would 
depend on the adiabaticity of antineutrino conversions
in the $H$ resonance layer, {\it i.e.} on the value of
$|U_{e3}|^2$. 

We shall consider the effects of neutrino
conversions in three possible regions of $|U_{e3}|^2$:
regions I, II and III as described in Sec.~\ref{adia}. 
When $|U_{e3}|^2$ lies in region III 
{\it i.e.} the $H$ resonance is inoperative, the resulting final 
spectra for a given solar neutrino solution are the same
as those obtained in Sec. \ref{sma}, \ref{lma} and \ref{vo}
for the normal mass hierarchy with the same mixing parameters.
Therefore, we need to consider only the cases when 
$|U_{e3}|^2$ lies in regions I or II.

%%%%%%%%%%%%%%%%%%%%%%%%%%%%%%%%%%%%%%%%%%%%%%%%%%%%%%%%%%%%%%%
\subsection{The scheme with the SMA solution}
\label{sma-inv}
%%%%%%%%%%%%%%%%%%%%%%%%%%%%%%%%%%%%%%%%%%%%%%%%%%%%%%%%%%%%%%%%%

The characteristics of the $\nu_e$ spectrum
are independent of the conversions at the $H$ resonance layer
[see eq.~(\ref{p-inv})].
The $L$ resonance is in the transition region so that  partial 
jumps from one level to another occur. As a consequence, from
the level crossing scheme (Fig.~\ref{generic}b), 
we get the following transitions:
$$
\nu_e \to \nu_1, \nu_2 ~~,~~
\nu_{\mu'} \to \nu_1, \nu_2 ~~,~~
\nu_{\tau'} \to \nu_3~~.
$$
Neglecting terms proportional to the small $|U_{e2}|^2$
in (\ref{p-inv}),
we get
$$p \approx P_L~~.$$
The $\nu_e$ spectrum is thus composite.

Let us now examine the final $\bar{\nu}_e$ spectrum.
Since the $L$ resonance is in the neutrino channel, 
the antineutrino transitions in the $L$ resonance layer are
adiabatic, so that $\bar{P}_L \approx 0$.
The characteristics of the final $\bar{\nu}_e$ spectrum
depend on the region in which the value of 
$|U_{e3}|^2$ lies.\\

{\bf Region I.}
The following transitions take place inside the star:
$$
\bar{\nu}_e \to \bar{\nu}_3~~,~~
\bar{\nu}_{\mu'} \to \bar{\nu}_2~~,~~
\bar{\nu}_{\tau'} \to \bar{\nu}_1~~.
$$
For $P_H \approx 0$ the Eq. (\ref{pbar-inv}) gives
$$
\bar{p} \approx |U_{e3}|^2 < 0.03~~.
$$
The final $\bar{\nu}_e$ spectrum is then almost completely
the original hard spectrum of $\nu_x$:
$$ F_{\bar{e}} \approx F_x^0 ~~. $$
The features of the final spectra are:
\begin{itemize}
\item The neutronization peak contains  both,
$\nu_e$ and $\nu_x$ fluxes.
\item The $\nu_e$ and $\nu_x$ spectrua are composite.
\item The $\bar{\nu}_e$ spectrum is practically
the original hard spectrum of $\nu_x$.
\end{itemize}
The earth matter effects on the $\nu_e$ spectrum
(\ref{fdiff-nu-inv}) are small: $\lsim 10 \%$
(see Sec.~\ref{sma}), but still may be observable.
The earth matter effects on the $\bar{\nu}_e$ spectrum
(\ref{fdiff-nubar-inv}) are negligible. They are 
suppressed by the factor of $\bar{P}_H \approx 0$
and further by the small mixing of the antineutrinos in 
the earth matter. \\

{\bf Region II.} Taking into account jumps of states 
at $H$ resonance we get the following transitions:  
$$
\bar{\nu}_e \to \bar{\nu}_1, \bar{\nu}_3~~,~~
\bar{\nu}_{\mu'} \to \bar{\nu}_2 ~~,~~
\bar{\nu}_{\tau'} \to \bar{\nu}_1, \bar{\nu}_3~~.
$$
Inserting $\bar{P}_L \approx 0$ in eq. (\ref{pbar-inv}), we find  
$$      
\bar{p} \approx \bar{P}_H \cos^2 \theta_\odot \approx \bar{P}_H~~.
$$
The following features are then expected in the final spectra:
\begin{itemize}

\item 
The neutronization peak contains both  $\nu_e$ and
$\nu_x$ fluxes.

\item 
All the three spectra, of $\nu_e$, $\bar{\nu}_e$ and 
$\nu_x$, are composite. In contrast with previous case 
(region I) now $\bar{\nu}_e$ spectrum is also composite. 

\end{itemize}
The earth matter effects 
(\ref{fdiff-nu-inv}) are small:  $\lsim 10 \%$
(see Sec.~\ref{sma}) for  $\nu_e$, and 
negligible for  the $\bar{\nu}_e$.\\

%%%%%%%%%%%%%%%%%%%%%%%%%%%%%%%%%%%%%%%%%%%%%%%%%%%%%%%%%%%%%%%%%%%%%%
\subsection{The scheme with the LMA solution}
\label{lma-inv}
%%%%%%%%%%%%%%%%%%%%%%%%%%%%%%%%%%%%%%%%%%%%%%%%%%%%%%%%%%%%%%%%%%%%%%%

The neutrino transitions at the  $L$ resonance are adiabatic, so that 
the following transitions take place according to the level crossing
scheme (Fig.~\ref{generic}d):
$$
\nu_e \to \nu_2 ~~,~~
\nu_{\mu'} \to \nu_1 ~~,~~
\nu_{\tau'} \to \nu_3~~.
$$
Inserting $P_L \approx 0$ in Eq. (\ref{p-inv}), we get
$$
p \approx |U_{e2}|^2 \approx \sin^2 \theta_\odot  \sim 0.2 - 0.4 ~~, 
$$
and consequently, the $\nu_e$ spectrum is  composite: 
$$ 
F_{e} \approx  \sin^2 \theta_\odot F_{e}^0 
+  \cos^2 \theta_\odot  F_x^0~~.   
$$

Let us examine the final $\bar{\nu}_e$ spectrum.
The antineutrino transitions in the $L$ resonance region
are also adiabatic, so that $\bar{P}_L \approx 0$.
Conversion in the $H$ resonance region, and therefore 
the characteristics of the final $\bar{\nu}_e$ spectrum
depend on  $|U_{e3}|^2$.\\

{\bf Region I.} 
Here the $H$ resonance is adiabatic and we get  
the following transitions: 
$$
\bar{\nu}_e \to \bar{\nu}_3~~,~~
\bar{\nu}_{\mu'} \to \bar{\nu}_2 ~~,~~
\bar{\nu}_{\tau'} \to \bar{\nu}_1~~.
$$
Since $P_H \approx 0$, the Eq. (\ref{pbar-inv}) gives
$$
\bar{p} \approx |U_{e3}|^2 < 0.03~~.
$$
and consequently, the  $\bar{\nu}_e$ spectrum is then almost completely
the original hard spectrum of $\nu_x$:
$$ 
F_{\bar{e}} \approx F_x^0 ~~. 
$$
The features of the final spectra are
the same as those of the scheme with 
the inverted mass hierarchy and
SMA solution (with $|U_{e3}|^2$ in region I).
The earth matter effects on the $\nu_e$ spectrum
(\ref{fdiff-nu-inv}) can be significant,
as can be seen from Fig.~\ref{earth-lma}a. 
In contrast, the earth  effects on the $\bar{\nu}_e$ spectrum
(\ref{fdiff-nubar-inv}) are negligible: 
they are suppressed by the factor of $\bar{P}_H \approx 0$ 
and  by the small mixing of the antineutrinos in 
the earth matter.\\

{\bf Region II.} The transitions in the $H$ resonance layer 
are  incomplete and 
therefore according to the level crossing scheme we have   
$$
\bar{\nu}_e \to \bar{\nu}_1, \bar{\nu}_3~~,~~
\bar{\nu}_{\mu'} \to \bar{\nu}_2 ~~,~~
\bar{\nu}_{\tau'} \to \bar{\nu}_1, \bar{\nu}_3~~.
$$
Inserting $\bar{P}_L \approx 0$ in eq. (\ref{pbar-inv}), we get 
$$ 
\bar{p} \approx \bar{P}_H |U_{e1}|^2   = 
\bar{P}_H \cos^2 \theta_\odot~.
$$
The features of the final spectra are
the same as those of the scheme with 
the inverted mass hierarchy and
SMA solution (with $|U_{e3}|^2$ in region II).
The earth matter effects can be  significant both 
in the $\nu_e$ 
(\ref{fdiff-nu-inv}) and in the $\bar{\nu}_e$ channels 
(\ref{fdiff-nubar-inv}) (see Fig.~\ref{earth-lma}).

%%%%%%%%%%%%%%%%%%%%%%%%%%%%%%%%%%%%%%%%%%%%%%%%%%%%%%%%%%%%%%%%%%%%%%
\subsection{The scheme with the VO solution}
\label{vo-inv}
%%%%%%%%%%%%%%%%%%%%%%%%%%%%%%%%%%%%%%%%%%%%%%%%%%%%%%%%%%%%%%%%%%%%%%

The conversions in the $H$ resonance layer 
does not influence $\nu_e$ flux. 
In the $L$ resonance layer the adiabaticity  is moderately broken:
$P_L \neq 0$.   
From the level crossing scheme (Fig.~\ref{generic}d),
the transitions taking place inside the star are:
$$
\nu_e \to \nu_1, \nu_2 ~~,~~
\nu_{\mu'} \to \nu_1, \nu_2 ~~,~~
\nu_{\tau'} \to \nu_3~~.
$$
From (\ref{p-inv}),
we get 
$$
\sin^2 \theta_\odot < p < \cos^2 \theta_\odot ~~.
$$
and since $\sin^2 \theta_\odot > 0.3$ we get 
$p = 0.3 - 0.7$. The $\nu_e$ spectrum is thus composite.

Let us now examine the final $\bar{\nu}_e$ spectrum.\\

{\bf Region I.} 
Due to adiabaticity in  the $H$ resonance layer, 
the following transitions take place:
$$
\bar{\nu}_e \to \bar{\nu}_3~~,~~
\bar{\nu}_{\mu'} \to \bar{\nu}_1, \bar{\nu}_2~~,~~
\bar{\nu}_{\tau'} \to \bar{\nu}_1, \bar{\nu}_2~~.
$$
Here $P_H \approx 0$, and from (\ref{pbar-inv}) we get
$$
\bar{p} \approx |U_{e3}|^2 < 0.03~~.
$$
The $\bar{\nu}_e$ spectrum is then almost completely
the original hard spectrum of $\nu_x$:
$$ F_{\bar{e}} \approx F_x^0 ~~. $$
The features of the final spectra are 
similar to those of the scheme with 
the inverted mass hierarchy and
SMA solution (with $|U_{e3}|^2$ in region I): 
the $\nu_e$ spectrum is composite whereas the  $\bar{\nu}_e$ 
spectrum is purely hard. 
The earth matter effects on both $\nu_e$ and $\bar{\nu}_e$
spectra are negligible. \\

{\bf Region II.} The adiabaticity is moderately broken both 
in the $H$ and in the $L$ resonance layers and  
the transitions proceed as 
$$
\bar{\nu}_e \to \bar{\nu}_1, \bar{\nu}_2, \bar{\nu}_3~~,~~
\bar{\nu}_{\mu'} \to \bar{\nu}_1, \bar{\nu}_2 ~~,~~
\bar{\nu}_{\tau'} \to \bar{\nu}_1, \bar{\nu}_2, \bar{\nu}_3~~.
$$
Correspondingly the Eq. (\ref{pbar-inv}) gives 
$$ \bar{p} \approx \bar{P}_H [\cos^2 \theta_\odot (1-\bar{P}_L)
+ \sin^2 \theta_\odot \bar{P}_L]~~.
$$
The features of the final spectra are
the same as those of the scheme with 
the inverted mass hierarchy and
SMA solution (with $|U_{e3}|^2$ in region II).
The earth matter effects on both $\nu_e$ and $\bar{\nu}_e$
spectra are negligible.

Let us summarize the results.  Some salient features of the final spectra
with
the inverted mass
hierarchy are the following:
\begin{itemize}

\item The neutronization peak contains  both $\nu_e$ and $\nu_x$ 
fluxes, in all the scenarios.

\item Also the final $\nu_e$ spectrum is composite in all
scenarios. The characteristics of this spectrum are independent
of the oscillation parameters in the $H$ resonance region.

\item The final $\nu_x$ spectrum is composite in all
scenarios.

\item When the value of $|U_{e3}|^2$ lies in region I, the final
$\bar{\nu}_e$ spectrum is almost the original hard spectrum of
$\nu_x$. For $|U_{e3}|^2$  in regions II and III the $\bar{\nu}_e$
spectrum is composite.

\item The earth matter effects on the $\nu_e$ spectrum can be 
significant for the scheme with the LMA solution, marginally
observable for the scheme with the SMA solution and 
negligible for the VO solution.
\item The earth matter effects on the $\bar{\nu}_e$ spectrum
can be significant only for the scheme with the LMA solution,
and when the antineutrino transitions at the $H$ resonance
are not adiabatic. In all the other scenarios, the earth matter
effects on $\bar{\nu}_e$ spectrum are negligible.
\item When the value of $|U_{e3}|^2$ lies in region III,
the final spectra (including the earth matter effects on them)
are indistinguishable from those obtained
with the scheme having the same oscillation parameters 
but the normal mass hierarchy.
\end{itemize}

\section{Signals of mixing and signatures of mixing schemes}
\label{signals}

The future detection of a supernova neutrino burst by the
underground neutrino detectors
has been discussed in \cite{totani, burrows, fuller, rrs}.
For a typical supernova at 10 kpc, about 5000 $\bar{\nu}_e$
events are expected to be detected at SK, and a few hundred
events each in SNO \cite{sno}, LVD \cite{lvd} and MACRO \cite{macro-sn}.
These detectors can reconstruct the energy of
the outgoing charged lepton from the $\bar{\nu}_e-p$
charged current interactions. SK and SNO can also observe
the direction of the charged lepton. In addition, SNO can
detect $\nu_e$ and reconstruct its energy through
the $\nu_e - d$ charged current interaction inside the
$D_2 O$.
The feasibility of measuring the absolute values
of the neutrino masses through the time-of-flight delays
has been studied
in \cite{stodolsky,vogel,choubey}.
Here, we concentrate on the features of the final
neutrino spectra that are relevant for identification of the
neutrino mass and flavor spectrum.

The effects of neutrino transitions on the 
final  neutrino spectra can be observed through, {\it e.g.}
(i) the partial or complete disappearance of the $\nu_e$
neutronization peak,
(ii) the apperance of hard / composite spectra of
$\nu_e$ or $\bar{\nu}_e$,
(iii) earth matter effects.
In this section, we shall elaborate on 
the effects that can be observed
in the earth detectors.

%%%%%%%%%%%%%%%%%%%%%%%%%%%%%%%%%%%%%%%%%%%%%%%%%%%%%%%%%%%%%%%%%
\subsection{The neutronization peak}
%%%%%%%%%%%%%%%%%%%%%%%%%%%%%%%%%%%%%%%%%%%%%%%%%%%%%%%%%%%%%%%%

The neutronization  peak can be identified as the burst  in the
neutrino signal 
during the first few milliseconds. 
If there is no neutrino mixing,
the neutrinos will be predominantly $\nu_e$. 
Let the expected ratio of the number of charged current (CC)
events to 
the number of neutral current (NC) events during the burst be 
$$
R_0^B \equiv \frac{N^{CC}}{N^{NC}}~~. 
$$ 
If neutrino conversions 
transform some $\nu_e$ into $\nu_x$,
the number of charged current events 
decreases while the number of neutral current
events remains unchanged. 
Then, the observed value of this ratio is
$R^B < R_0^B$.

When the value of $|U_{e3}|^2$ lies in the region I
in the case of the normal mass hierarchy
(so that the neutrino transitions in the $H$ resonance
layer are adiabatic),
the $\nu_e$ flux, and hence the number of 
charged current events,
gets suppressed by a factor of $\approx |U_{e3}|^2$ (see
Secs. \ref{sma}, \ref{lma}, \ref{vo}):
\beq
\frac{R^B}{R_0^B} \approx |U_{e3}|^2 \leq 0.03~~.
\eeq
The observation of $R_B > 0.03~ R_0^B$ would then point
against $|U_{e3}|^2$ in region I for the normal mass
hierarchy.
According to  (\ref{reg1limit}),
this will allow us to put
an upper bound on the mixing parameter $|U_{e3}|^2$:
\beq
|U_{e3}|^2 \lsim 3 \cdot 10^{-4}~~.
\label{ue3bd}
\eeq

%%%%%%%%%%%%%%%%%%%%%%%%%%%%%%%%%%%%%%%%%%%%%%%%%%%%%%%%%%%%%%%%%%%%%%%%%
\subsection{The $\langle E \rangle$ inequalities  and high energy
``tails''}
\label{compare}
%%%%%%%%%%%%%%%%%%%%%%%%%%%%%%%%%%%%%%%%%%%%%%%%%%%%%%%%%%%%%%%%%%%%%%%%%%

In the absence of mixing, one should observe 
$\langle E_{\nu_e} \rangle < \langle E_{\bar{\nu}_e} \rangle$.
The inversion of this inequality, {\it i.e.} 
$\langle E_{\nu_e} \rangle > \langle E_{\bar{\nu}_e} \rangle$
is the signature of the $\nu$-conversion. It  
implies  that the contribution of the converted original hard $\nu_x$
spectrum to the final $\nu_e$ flux is significantly
larger  than its contribution to the final $\bar{\nu}_e$ flux.

In the case of the normal mass hierarchy, the
inverted inequality for $\langle E \rangle$  will be observed  when the 
transitions in the $H$ resonance layer are adiabatic.
The observation of the normal inequality 
$\langle E_{\nu_e} \rangle < \langle E_{\bar{\nu}_e} \rangle$
can  exclude this scenario.

In the case of inverted mass hierarchy, the adiabatic  transitions in the
$H$ resonance layer preserve  the original inequalities for 
$\langle E \rangle$: 
$\langle E_{\nu_e} \rangle < \langle E_{\bar{\nu}_e} \rangle$.
Therefore, an observation of the inequality
$\langle E_{\nu_e} \rangle > \langle E_{\bar{\nu}_e} \rangle$
rules out the scenarios which have
the inverted mass hierarchy with the value of
$|U_{e3}|^2$  in region I.
If the mass hierarchy is known to be the inverted one,
an upper bound (\ref{ue3bd}) on the mixing parameter $|U_{e3}|^2$
can be  obtained.

The relative difference of the $\nu_e, \bar{\nu}_e$ and 
$\nu_x$ spectra is especially significant in  
the high energy ends (``tails'') of the spectra, 
where the fluxes decrease exponentially with the
increase of energy. In the absence of mixing, one expects
\beq
\frac{N_e(E>E_{tail})}{N_{\bar{e}}(E>E_{tail})} \ll 1, ~~
\frac{N_{\bar{e}}(E>E_{tail})}{N_x(E>E_{tail})} \ll 1
\eeq
for high enough energy $E_{tail}$ (energy greater than the
peak energies of the spectra). 
Here $N_i(E>E_{tail})$ is the integrated
number of $\nu_i-N$ charged current events
above the energy $E_{tail}$.

The observation of the high energy ends of the $\nu_e$ and 
$\bar{\nu}_e$ spectra,  where the contributions due to the
original
$\nu_e$ and $\bar{\nu}_e$ fluxes would be negligible,
allows one to  measure  the contribution of the
original hard $\nu_x$ spectrum to the final
$\nu_e$ and $\bar{\nu}_e$ spectra.

Important conclusions can be drawn from the studies
of the ratio
\beq
R(E_{tail}) \equiv \frac{N_{e}(E>E_{tail})}
{N_{\bar{e}}(E>E_{tail})}
\eeq
as $E_{tail}$ increases. 

1) 
$R(E_{tail}) \rightarrow \infty$:

This indicates that the tail of the $\nu_e$ spectrum extends to
significantly higher energies than that of the
$\bar{\nu}_e$ spectrum.
This will testify for $\nu_e \leftrightarrow \nu_x$ conversions.
Moreover,
it will be  a clear signal that the
$\bar{\nu}_e \leftrightarrow \nu_x$ conversions
are negligible, {\it i.e.} the $\bar{\nu}_e$ spectrum
is the original soft spectrum.
This would be an indication of
(i) the SMA solution in the conventional
hierarchy (sec.~\ref{sma}), or
(ii) highly non-adiabatic $H$ transitions with the SMA
solution in the inverted hierarchy (sec. \ref{sma-inv}).
If the mass hierarchy is known to be the inverted one,
then this signal indicates that the value of $|U_{e3}|^2$
lies in region III. From (\ref{reg2limit}), we obtain
a very strong bound on the parameter  $|U_{e3}|^2$:
\beq
4 |U_{e3}|^2 \lsim 10^{-5}~~.
\label{strong-bound}
\eeq

2) $R(E_{tail}) \rightarrow 0$:

Notice that the limit $R(E_{tail}) \rightarrow 0$  would be true 
in the absense of any neutrino conversion at all, or 
in the absence of conversion of the electron neutrinos. 
This situation is not realized in any of the scenarios  
we have discussed. 
The observation of $R(E_{tail}) \rightarrow 0$ would 
exclude all the scenarios under consideration and thus testify for 
the solution of the solar neutrino problem which differs from 
the MSW or VO  solutions. 
Another possibility could be related to a compact star with 
a very sharp density profile, so that the $L$ resonance is in the 
non-adiabatic region. In this case, the solar solution 
has to be SMA and the $H$ resonance has to be
inoperative for neutrinos.

3) 
$R(E_{tail}) \rightarrow const$:

This would testify for the contribution of the original $\nu_x$ 
to both final $\nu_e$ and $\bar{\nu}_e$ spectra.
Moreover, the value of the
constant will give us information about 
the mixing parameters. \\

In principle, the spectra of $\nu_x$ can be reconstructed
by having a series of neutral current detectors with different
thresholds. 
The $\nu_x$ spectrum
will always have a dominating hard component.
Therefore, the comparison of the high energy tails
of the $\nu_e, \bar{\nu}_e$ with that of $\nu_x$
can directly give the measure of the hard component of
the $\nu_e, \bar{\nu}_e$ spectra.
The ability to measure the $\nu_x$ spectrum would also be
useful in order to check for the presence of sterile neutrinos
\cite{ds-sterile}.

%%%%%%%%%%%%%%%%%%%%%%%%%%%%%%%%%%%%%%%%%%%%%%%%%%%%%%%%%%%%%%%
\subsection{The identification of a composite spectrum}
%%%%%%%%%%%%%%%%%%%%%%%%%%%%%%%%%%%%%%%%%%%%%%%%%%%%%%%%%%%%%%

The final $\nu_e$ ($\bar{\nu}_e$) spectrum  can be qualitatively
divided into  three types: (a) the original ``soft'' spectrum
of the $\nu_e$ ($\bar{\nu}_e$) [corresponding to the
survival probability $p = 1$ ($\bar{p} = 1$)
which would be the case in the absence of conversion],
(b) the ``hard'' original spectrum of $\nu_x$
[corresponding to the
survival probability $p =0$ ($\bar{p} = 0$)
which would be the case when there is a complete interchange
of spectra],
and (c) the ``composite'' spectrum, which is a mixture of the
original soft and the hard spectra in comparable proportions.

In order to distinguish between (a) and (b), the values of
``temperatures'' (or average energies) of the original
spectra need to be known, or the $\nu_e$ and $\bar{\nu}_e$
spectra have to be compared (as in Sec.~\ref{compare}).
A  composite spectrum may be identified 
without the knowledge of the initial temperatures 
and independently of the observation of the other spectrum,
through the ``broadening'' phenomenon.

\begin{figure}[htb]
\hbox to \hsize{\hfil\epsfxsize=10cm\epsfbox{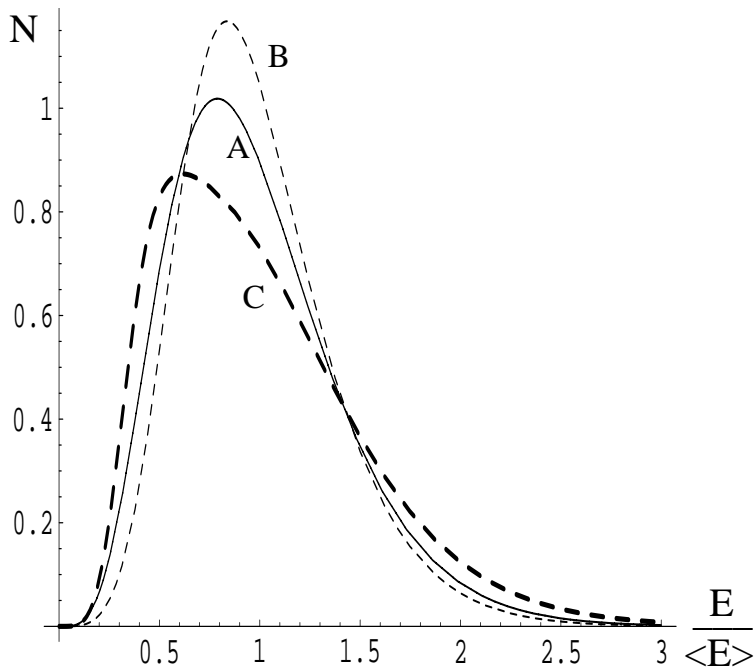}\hfil}
\caption{~~The number of $\nu-N$ charged current events for
(A) a Fermi-Dirac spectrum ($T=3,\eta=0$): the solid line,
(B) a pinched spectrum ($T=3,\eta=3$): the dotted line, and
(C) a mixture of two pinched spectra ($T=3,\eta=3$
and $T=8,\eta=1$ with $E_{na}=5$ MeV): the dashed line.
\label{broad-abc}}
\end{figure}

As described in Sec.~\ref{fluxes},
the instantaneous original neutrino spectra   
have a narrower energy distribution than the Fermi-Dirac one
(see Fig.~\ref{pinched}).
The mixing between two 
neutrino species with different mean energies
gives rise to an effective broadening of the spectrum
(see Fig.~\ref{broad-abc}), so that
The final spectrum need not be pinched even though the
two individual original spectra were pinched.
In other words,
the effective $\eta$ for a mixed spectrum may be
negative even though the individual $\eta_i$'s of the
constituent spectra were positive.
The broadening of the spectrum can be checked
 by fitting the final spectrum with
the parameters $T_i$ and $\eta_i$ 
(\ref{spectrum}) and establishing the sign of 
$\eta_i$.

Note that the ``pinching'' phenomenon is established
in a model independent manner only 
for a constant mean energy of the spectrum.
A considerable variation in the average energy 
of neutrinos of a 
given species during the cooling phase could be 
responsible for the broadening of the 
time integrated spectrum 
even in the absence of mixing. 
Therefore, in order to establish the broadening due to mixing,
it is crucial to 
study the spectra in short intervals of time.

Clearly, it will be difficult to observe broadening of 
spectra,  
if the mean energies of the original spectra are very close,
or if one of the original spectrum dominates in the
final mixed spectrum.

The observation of broadening of the $\nu_e$ 
spectrum would point against 
an adiabatic $H$ resonance in the normal hierarchy
(see secs.~\ref{sma}, \ref{lma}, \ref{vo}).
This would imply 
a strong upper bound (\ref{ue3bd}) on the value of
$|U_{e3}|^2$.

The observation of broadening of the $\bar{\nu}_e$ 
spectrum would be a strong indication against
\begin{itemize}
\item the SMA solution with the normal hierarchy
(Sec.~\ref{sma}),
\item completely adiabatic $H$ transitions in the case of
inverted hierarchy (secs.~\ref{sma-inv}, \ref{lma-inv},
\ref{vo-inv}), and 
\item highly non-adiabatic $H$-transitions for the SMA solution 
with the inverted hierarchy (Sec.~\ref{sma-inv}).
\end{itemize}
In particular, if the mass hierarchy is inverted, 
the broadening of $\bar{\nu}_e$ spectrum indicates 
that the $H$ resonance is not completely adiabatic.
This gives a strong upper bound (\ref{ue3bd})
on $|U_{e3}|^2$.

If both the $\nu_e$ and $\bar{\nu}_e$ spectra are established
to be composite, then the upper bound  (\ref{ue3bd})
on $|U_{e3}|^2$ holds irrespective of the type of
mass hierarchy.

%%%%%%%%%%%%%%%%%%%%%%%%%%%%%%%%%%%%%%%%%%%%%%%%%%%%%%%%%%%%%%%%%%%%%%
\subsection{Earth matter effects}
%%%%%%%%%%%%%%%%%%%%%%%%%%%%%%%%%%%%%%%%%%%%%%%%%%%%%%%%%%%%%%%%%%%%%%

The earth matter effects 
can be observed through the comparison between the spectra at two
detectors, or through the studies of the distortion of 
a spectrum in one detector (See Figs.~\ref{earth-sma},
~\ref{earth-lma}).

The earth matter effects are
significant 
in the neutrino channel only in
the cases of
(i) normal mass hierarchy with the SMA or LMA scenario 
and a non-adiabatic $H$ resonance, 
(ii) inverted mass hierarchy with the SMA or LMA scenario.
In the antineutrino channels, significant earth
matter effects are observed (sec.~\ref{earth-antin}) only with
(i) normal mass hierarchy with the LMA scenario, 
(ii) inverted mass hierarchy with LMA scenario and a
non-adiabatic $H$ resonance.

\section{Identification of the mass spectrum}
\label{concl}

We show that with the $3\nu$ schemes
which explain the solar and atmospheric  neutrino data,
one can make rather reliable predictions of the
conversion effects for supernova neutrinos.
The predictions differ for different schemes,
which opens up the possibility of discriminating among them
using the data from the neutrino bursts.

We have studied the manifestation of the conversion effects
in
(i) The flavor composition of the neutronization peak:
the main effect here is the partial or complete change of the flavor
of the peak.
This can be established  by the comparison of the fluxes
detected by charged current and neutral current interactions
during the neutronization burst.
(ii) Modifications of the ${\nu}_e$ and $\bar{\nu}_e$ spectra:
Here one expects an appearance of hard or composite
${\nu}_e$ and/or $\bar{\nu}_e$ spectra due to mixing, 
instead of their original soft spectra.
One can identify these effects of mixing by studying
the average energies of spectra, the tails of spectra,
and searching for the widening of the spectra.
(iii) Earth matter effects: This can be done
by detailed studies of the shapes of the energy spectra, or
by the comparison of signals observed in different detectors.

The final spectra of $\nu_e$ and $\bar{\nu}_e$ can be 
characterized by the values of the survival probabilities
$p$ and $\bar{p}$ respectively. 
The conversion probabilities depend significantly on the 
value of $|U_{e3}|^2$. We have divided the whole range of possible 
values of $|U_{e3}|^2$ in three regions: I, II, and III, and
have made definite predictions in each of these regions. 
Table~\ref{ppbar} gives
the values of the survival probabilities for the neutrino
mass spectra under discussion.

\begin{table}
\begin{center}
\begin{tabular}{|llr|c|c|}
\hline
 & & & $p$ & $\bar{p}$ \\
\hline
I & SMA & normal & $|U_{e3}|^2$ & 1 \\
& & inverted & $P_L$ & $|U_{e3}|^2$ \\
 & & & & \\
& LMA & normal & $|U_{e3}|^2$ & $\cos^2 \theta_\odot$ \\
& & inverted & $\sin^2 \theta_\odot$ & $|U_{e3}|^2$ \\ 
 & & & & \\
& VO & normal & $|U_{e3}|^2$ & $[\sin^2 \theta_\odot, 
\cos^2 \theta_\odot]$ \\
&  & inverted & $[\sin^2 \theta_\odot, 
\cos^2 \theta_\odot]$ &  $|U_{e3}|^2$ \\
\hline
II & SMA & normal & $[|U_{e3}|^2, P_L]$ & 1 \\
& & inverted & $P_L$ & $\bar{P}_H$ \\
 & & & & \\
& LMA & normal & $\sin^2 \theta_\odot P_H$ & $\cos^2 \theta_\odot$ \\
& & inverted & $\sin^2 \theta_\odot$ & $\cos^2 \theta_\odot \bar{P}_H$ \\
 & & & & \\
& VO & normal & $[|U_{e3}|^2, \cos^2 \theta_\odot]$ &
$[\sin^2 \theta_\odot, \cos^2 \theta_\odot]$ \\
&  & inverted & $[\sin^2 \theta_\odot, \cos^2 \theta_\odot]$ &
$[\sin^2 \theta_\odot \bar{P}_H, \cos^2 \theta_\odot \bar{P}_H]$ \\
\hline
III & SMA & normal & $P_L$ & 1 \\
& & inverted & $P_L$ & 1 \\
 & & & & \\
& LMA & normal & $\sin^2 \theta_\odot$ & $\cos^2 \theta_\odot$ \\
& & inverted & $\sin^2 \theta_\odot$ & $\cos^2 \theta_\odot$ \\
 & & & & \\
& VO & normal & $[\sin^2 \theta_\odot, \cos^2 \theta_\odot]$ &
$[\sin^2 \theta_\odot, \cos^2 \theta_\odot]$ \\
&  & inverted & $[\sin^2 \theta_\odot, \cos^2 \theta_\odot]$ &
$[\sin^2 \theta_\odot, \cos^2 \theta_\odot]$ \\
\hline
\end{tabular}
\end{center}
\caption{The values of the survival probabilities $p$ and $\bar{p}$
for various scenarios. $[x,y]$ indicates that the value of the
survival probability lies between $x$ and $y$. 
\label{ppbar}}
\end{table}

\begin{table}
\begin{center}
\begin{tabular}{|llr|c|cc|cc|}
\hline
 & & & Neutronization & \multicolumn{2}{c|}{spectrum} & \multicolumn{2}{c|}{Earth effects}\\
&&  & peak &  $\nu_e$ & $\bar{\nu}_e$ &
$\nu_e$ & $\bar{\nu}_e$ \\ 
\hline
I & SMA & normal & $\approx \nu_x$ & hard & soft & $\approx 0$ & $\approx 0$ \\
& & inverted & $\nu_e ,~\nu_x$ & composite & hard & $\surd$ & $\approx 0$
 \\
 & & & & & & & \\
& LMA & normal & $\approx \nu_x$ & hard & composite & $\approx 0$ & $\surd$ \\
& & inverted & $\nu_e ,~\nu_x$ & composite & hard & $\surd$ & $\approx 0$ \\
 & & & & & & & \\
& VO & normal & $\approx \nu_x$ & hard & composite & $\approx 0$ & $\approx 0$ \\
&  & inverted & $\nu_e ,~\nu_x$ & composite & hard & $\approx 0$ &  $\approx 0$\\
\hline
II & SMA & normal & $\nu_e ,~\nu_x$ & composite & soft & $\surd$ & $\approx 0$ \\
& & inverted & $\nu_e ,~\nu_x$ & composite & composite &  $\surd$ & $\approx 0$ \\
 & & & & & & & \\
& LMA & normal &$\nu_e ,~\nu_x$  & composite & composite & $\surd$ & $\surd$\\
& & inverted & $\nu_e ,~\nu_x$ & composite & composite & $\surd$ & $\surd$\\
 & & & & & & & \\
& VO & normal & $\nu_e ,~\nu_x$ & composite & composite & $\approx 0$ &  $\approx 0$\\
&  & inverted & $\nu_e ,~\nu_x$ & composite & composite &  $\approx 0$ & $\approx 0$\\
\hline
III & SMA & normal & $\nu_e ,~\nu_x$  & composite & soft & $\surd$ &  $\approx 0$\\
& & inverted & $\nu_e ,~\nu_x$  & composite & soft & $\surd$ & $\approx 0$ \\
 & & & & & & & \\
& LMA & normal & $\nu_e ,~\nu_x$ & composite & composite & $\surd$ & $\surd$\\
& & inverted & $\nu_e ,~\nu_x$ & composite & composite & $\surd$ & $\surd$\\ 
 & & & & & & & \\
& VO & normal &  $\nu_e ,~\nu_x$ & composite & composite & $\approx 0$ & $\approx 0$\\
&  & inverted &  $\nu_e ,~\nu_x$ & composite & composite & $\approx 0$ & $\approx 0$\\
\hline
\end{tabular}
\end{center}
\caption{The dependence of the final spectra on (1) the solution
of the solar neutrino problem (2) the nature of hierarchy (normal
or inverted) and (3) the value of $|U_{e3}|^2$. ``Soft'' in the
$\nu_e$ column refers to the original $\nu_e$ spectrum and
``soft'' in the $\bar{\nu}_e$ column refers to the original 
$\bar{\nu}_e$ spectrum. ``Hard'' refers to the original
$\nu_x$ spectrum. ``Composite'' implies that the final
spectrum has both the soft and the hard components.  
In the Earth Matter
Effects columns, a $\surd$ indicates the possibility of
significant earth matter effects.
\label{softhard}}
\end{table}

The qualitative features of the final neutrino fluxes 
for various neutrino mass spectra, 
obtained in sec.~\ref{final-spec}-\ref{inverted},
are summarized in
Table~\ref{softhard}. 
Let us first make some general observations
from the table.

1. The complete disappearance of the $\nu_e$ neutronization peak 
(strong change of its flavor) and a pure hard spectrum of $\nu_e$ 
during the cooling stage always appear together. Their observation
will testify for $|U_{e3}|^2$ in the region I and normal mass hierarchy,
irrespective of the solution of the solar neutrino problem.

2. A pure hard $\bar{\nu}_e$ spectrum is the signature of the inverted
mass hierarchy and  $|U_{e3}|^2$ in the region I, irrespective of the
solar neutrino solution. 

3. A soft $\bar{\nu}_e$ spectrum can appear only in the schemes with
the SMA solution of the solar neutrino problem.

4. The observation of any earth matter effects will rule out
the scenarios with the VO solution.

5. The earth matter effects in $\bar{\nu}_e$ spectrum will be 
the signature of the LMA solution.

6. If the earth matter effects are observed in the
neutrino channels but not in the antineutrino
channels, we either have 
the normal mass hierarchy with $|U_{e3}|^2$ in
regions II or III, or the inverted mass hierarchy.

7. If the earth matter effects are significant in both the 
$\nu_e$ and $\bar{\nu}_e$ channels, the solar neutrino solution is
LMA and $|U_{e3}|^2$ lies in regions II or III. \\

Let us now systematically consider all the possible
combinations of the observations.
In general, the ``observations'' consist of the following 
features of the spectra:
\begin{itemize}
\item The flavor of the neutronization peak can be
either mixed ($\nu_e$, $\nu_x$) or (almost) completely changed
($\nu_x$). 
\item  The $\nu_e$ spectrum at the cooling stage can be either hard
or composite. 
\item The $\bar{\nu}_e$ spectrum can be of all possible types:
soft (unchanged), hard, composite.
\item The earth matter effects can be absent completely,  
or observed in one of channels ($\nu_e$ or $\bar{\nu}_e$), 
or observed in both channels.
\end{itemize}

The following conclusions can be drawn from the
combinations of the above observations:

(A) Completely changed flavor ($\nu_x$) of the neutronization peak,
hard $\nu_e$ spectrum, and soft $\bar{\nu}_e$ spectrum.
This configuration can be realized in only one case:
normal hierarchy, SMA solution of the $\nu_{\odot}$-problem
and $|U_{e3}|^2$ in the region I. 

(B) Pure $\nu_x$ - neutronization peak, hard $\nu_e$ spectrum,
and composite $\bar{\nu}_e$ spectrum. This implies
normal mass hierarchy, $|U_{e3}|^2$ in the region I, and
two possible $\nu_\odot$ solutions, LMA or VO.
The observation of earth matter effects can distinguish between
these two possibilities. For the LMA solution, 
the earth effects will be present in the
$\bar{\nu}_e$ channel and absent in the $\nu_e$ channel.

(C) Mixed ($\nu_e$, $\nu_x$) flavor neutronization peak,
composite $\nu_e$ spectrum, and soft $\bar{\nu}_e$ spectrum.
This configuration can be realized in the three
different cases: 
(i) SMA solution, normal mass hierarchy,
and $|U_{e3}|^2$ in the region II, 
(ii) SMA solution, normal mass hierarchy, 
but $|U_{e3}|^2$ in the region III,
(iii) SMA solution, inverted hierarchy, $|U_{e3}|^2$ in the
region III. 
Clearly such a configuration is the signature of the SMA
solution of the $\nu_{\odot}$-problem. 
Earth matter effect is
expected to be weaker in the case (i).

(D) Mixed ($\nu_e$, $\nu_x$) flavor  neutronization peak,
composite $\nu_e$ spectrum,  composite  $\bar{\nu}_e$ spectrum.
This combination of observations can be realized in
many  cases:
(i) SMA solution, inverted hierarchy, $|U_{e3}|^2$ in the region II,
(ii-v) LMA solution, normal or inverted hierarchy, 
$|U_{e3}|^2$ in the regions II or III,
(vi-ix) VO solution, normal or inverted hierarchy, 
$|U_{e3}|^2$ in the regions II or III.
Studies of the earth effects can give an important  
criterion for discriminating between the above cases:
they are absent in the VO cases (vi-ix) and should be observable 
in the $\nu_e$ channel but not in $\bar{\nu}_e$ channel
in the SMA  case (i).
In the LMA case, the earth effects are expected both in
the $\nu_e$ and $\bar{\nu}_e$ channels.
Here a further discrimination  can be done between (ii-v)
by studying the degree of compositness: 
in the case of normal hierarchy one expects
stronger effects in the neutrino channel.

(E) Mixed ($\nu_e$, $\nu_x$) flavor neutronization peak,
composite $\nu_e$ spectrum,  hard  $\bar{\nu}_e$ spectrum.
Such a combination of observations is realized with
inverted  hierarchy, $|U_{e3}|^2$ in the region I, and
SMA or LMA or VO solution. That is, the combination does
not depend on the solution of the
$\nu_{\odot}$-problem and
turns out to be the signature of the inverted mass hierarchy.
In the cases with  SMA or LMA solutions the earth effect can be
observed in the $\nu_e$ channel only. No earth effect
is expected in the VO case.

No other combination of the observables is realized in the
mixing scenarios under disussion. \\

Establishing the region in which $|U_{e3}|^2$ lies is 
equivalant to giving 
bounds on the value of this mixing parameter, which correspond
to the borders of the region.
The borders depend on the precise 
density profile of the progenitor and can vary within 
a factor of 2 - 3. Thus, from SN data we can get bounds on $|U_{e3}|^2$ 
with an uncertainty of a factor of 2 - 3.
In principle the value of $|U_{e3}|^2$ can also be measured 
(again within a factor of 2 -3) in 
the transition region where the effect (survival probability)  
depends  substantially on $U_{e3}$. 

The observations
(A), (B), and (E) above establish $|U_{e3}|^2$ in region I, thus 
giving a lower bound of $|U_{e3}|^2 \gsim 3 \cdot 10^{-4}$. 
The observations
(C) and (D) establish $|U_{e3}|^2$ to be in region II or III,
thus giving an upper bound of $|U_{e3}|^2 \lsim 3 \cdot 10^{-4}$.
The upper bound obtained in the case of the observations
(C) and (D) is more than an order of magnitude stronger than the
one the long baseline experiments are planning to achieve.
The observation of the broadening of both the 
$\nu_e$ and $\bar{\nu}_e$ spectra, or
significant earth matter effects on both the 
$\nu_e$ and $\bar{\nu}_e$ spectra are sufficient to
establish this bound. \\

A number of possible observations 
can rule out all the scenarios under discussion.
The observation of
(a) a pure $\nu_e$ neutronization peak and (b) a soft
$\nu_e$ spectrum during the cooling stage is not possible
with any of these neutrino mass spectra. 
The pairs of observations
(a) pure $\nu_x$ neutronization peak and a composite $\nu_e$ spectrum,
(b) mixed flavour neutronization peak and a pure hard $\nu_e$
spectrum, and
(c) a hard spectrum for both $\nu_e$ and $\bar{\nu_e}$
are also fatal for all the scenarios.

The final neutrino spectra can thus help in resolving three 
kinds of ambiguities that remain to be resolved
with the current data:
(i) the  solution of the solar neutrino problem,
(ii) the type of mass hierarchy (sign of $\Delta m^2$), and
(iii) the  value of $|U_{e3}|^2$. \\

The implications of results of this paper will depend
on when  neutrino burst from the Galactic supernova will be detected.

A number of results can be obtained from 
other experiments unrelated to the supernova neutrinos.
Clearly, all the schemes considered above will
be rejected if sterile
neutrinos will be discovered in the oscillations of
solar or atmospheric neutrinos, or if the LSND result will
be confirmed.
In this case, one will be forced to consider the schemes
with sterile neutrinos \cite{ds-sterile}.

On the other hand, there is a good chance that
within several years the existing and future experiments will allow
us to identify one of solutions of the solar 
neutrino problem considered in this paper.
This will significantly diminish the number of possible schemes
and will allow us to further sharpen the predictions 
of the effects for supernova neutrinos.

The purpose of this paper was to indicate the effects
which can in principle testify
for neutrino conversions
and various features of the neutrino mass and flavor spectra.
Clearly, further detailed studies are needed
for specific detectors to clarify the detectability of
the discussed effects and to
conclude how far we can go
in the program of identification of the neutrino mass spectrum.

\section*{Acknowledgements}
One of us (A. D.) would like to thank J. Lattimer and G. Raffelt for
discussions.

\end{document}